\documentclass[aps,pre,reprint,twocolumn,floats,floatfix,superscriptaddress]{revtex4-2}

\usepackage{blindtext}  % This is used to generate meaningless text; you can remove this line
\usepackage{graphicx}
\usepackage{amsmath}
\usepackage{amsfonts}
\usepackage{amssymb}
\usepackage{amstext}
\usepackage[english]{babel}
\usepackage{helvet}
\usepackage{microtype}
\usepackage[pdftex,hidelinks]{hyperref}
\hypersetup{
    colorlinks=true,
    linkcolor=blue,
    filecolor=blue,
    urlcolor=blue,
    citecolor = blue,
}
\usepackage{enumitem}
\usepackage{xcolor}
\usepackage[export]{adjustbox}
\usepackage[caption=false]{subfig}

\graphicspath{{images/}}

\newcommand{\ket}[1]{|#1\rangle}

\DeclareMathOperator{\tr}{Tr}

\begin{document}

% \title{Is the entanglement-entropy area law an impediment to real-space renormalization group in 3D?}
\title{
    Three-dimensional real space renormalization group with well-controlled approximations
}


\author{Xinliang Lyu}
\email[]{xlyu@ihes.fr}
%\thanks{}
\altaffiliation{Present address:
    Institut des Hautes \'Etudes Scientifiques,
    91440 Bures-sur-Yvette, France
}
\affiliation{
    Institute for Solid State Physics, The University of Tokyo,
    Kashiwa, Chiba 277-8581, Japan
}
% Last author
\author{Naoki Kawashima}
\email[]{kawashima@issp.u-tokyo.ac.jp} 
\affiliation{
    Institute for Solid State Physics, The University of Tokyo, 
    Kashiwa, Chiba 277-8581, Japan
}
\affiliation{
    Trans-scale Quantum Science Institute, The University of Tokyo
    7-3-1, Hongo, Tokyo 113-0033, Japan
}

\date{March 21, 2025}
%\date{\today}

\begin{abstract}
We make Kadanoff’s block idea into a reliable three-dimensional (3D) real space renormalization group (RG) method.
Kadanoff’s idea, expressed in spin representation, offers a qualitative intuition for clarifying scaling behavior in criticality, but has difficulty as a quantitative tool due to uncontrolled approximations.
A tensor-network reformulation equips the block idea with a measure of RG errors.
In 3D, we propose an entanglement filtering scheme to enhance such a block-tensor map, with the lattice reflection symmetry exploited.
When the proposed RG is applied to the cubic-lattice Ising model, the RG errors are reduced to about 2\% by retaining more couplings.
The estimated scaling dimensions of the two relevant fields have errors 0.4\% and 0.1\% in the best case, compared with the accepted values.
The proposed RG is promising as a systematically-improvable real space RG method in 3D.
The unique feature of our method is its ability to numerically obtain a 3D critical fixed point in a high-dimensional tensor space.
A fixed-point tensor contains much more information than a handful of observables estimated in conventional techniques for analyzing critical systems.
\end{abstract}

\maketitle

\newpage

% Text body starting here
% Introduction
\section{Introduction}
The renormalization group (RG) idea offers both a theoretical framework for understanding the universality in critical phenomena~\cite{Wilson:1971a} and a practical approach for calculating the critical exponents~\cite{Wilson:1971b,Wilson:1972} that quantify a universality class.
The key operation of the RG idea is a coarse-graining process to generate a series of descriptions of a system in increasingly larger length scales, represented by a flow in the coupling-constant space known as an RG flow.
Phase transitions belonging to the same universality class correspond to the same critical fixed point of these RG flows.
The critical exponents can be calculated from scaling dimensions $\{x_i\}$, which are defined using the eigenvalues $\{\lambda_i\}$ of the linearized RG flow around the critical fixed point,
\begin{align}
    \label{eq:intro:rg2scaleD}
    b^{d - x_i} = \lambda_i,
\end{align}
where $d$ is the spatial dimensionality of the thermodynamic system and $b$ is the rescaling factor of the coarse-graining.

Kadanoff’s block-spin idea~\cite{Kadanoff:1966} is the prototype of the RG in real space:
a block of spins are replaced with one coarser spin, provided that the partition function is preserved.
For spatial dimensionality $D \geq 2$, if the partition function is preserved exactly, a single step of this RG map entails all possible interaction terms among spins with all ranges~\cite{Kardar:2007}.
When designing a practical numerical RG method, we have to truncate the coupling-constant space down to finite dimensions.

%{\color{orange}
As an example, we focus on a well-known truncation scheme for the block-spin idea.
%}
Using a drastic approximation, Migdal~\cite{Migdal:1975a} and Kadanoff~\cite{Kadanoff:1976} proposed a bond-moving scheme that decouples the interactions in different spatial directions.
This Migdal-Kadanoff (MK) approach maintains the nearest-neighbor interaction form and is applicable in any spatial dimensionality.
However, MK approach only works well when the transition temperature is close or equal to zero, or equivalently the spatial dimensionality is near the lower critical dimension.
This is, for example, not true for the Ising model in 2D and 3D, and the estimate of thermal exponents $\nu$ has error almost $50\%$ for 3D and $25\%$ for 2D, which are very crude approximations.
Martinelli and Parisi~\cite{Martinelli:Parisi:1981} designed a systematical improvement of MK approach.
For the 3D Ising model, the first-order corrections move MK estimates of $\nu$ closer to the accepted value.
However, no higher-order corrections have been computed.

Above all, there is no \emph{a priori} quantitative metric to assess the approximations made in these spin-based  RG transformations; the justification usually comes from an \emph{a posteriori} comparison with the accepted estimates of various critical exponents.
Therefore, apart from an intuitive feeling that they could work, one might feel uneasy to trust them as precise numerical methods~\cite{Kadanoff:2000:book}.

In this paper, we propose an RG transformation in real space for any three-dimensional (3D) classical statistical system on a lattice~\footnote{
    With some modifications, it can also be applied to any (2+1)D quantum system.
}.
Our method is based on a tensor-network reformulation of the real space RG~\cite{Levin:Nave:2007,Evenbly:TNRalgo}, which has a natural metric for quantifying the RG approximation errors.
Our main contribution is designing a workable 3D entanglement filtering (EF) scheme based on Refs.~\cite{Hauru:Delcamp:Mizera:2018,Evenbly:2018} to enhance the usual block-tensor transformation.
%{\color{orange}
Lattice reflection symmetry is exploited to simplify the EF process.
%}
An EF process is essential for obtaining a critical fixed-point tensor and taming the growth of RG errors with the RG step in 3D.
%{\color{orange}
    As a benchmark, we apply the proposed RG to the cubic-lattice Ising model.
%}
The estimated scaling dimensions of the spin and the energy density fields $x_{\sigma}, x_{\epsilon}$, compared with the conformal bootstrap estimates~\cite{Kos:Poland:Simmons:2016,Chang:Dommes:Erramilli:2024}, have relative errors $0.4\%$ and $0.1\%$ in the best case.

% Block-tensor map
\section{Block-tensor transformation}
A reformulation of the real space RG in the tensor-network language equips it with a natural measure of the RG approximation error~\cite{Levin:Nave:2007,Evenbly:TNRalgo}.
This reformulation is inspired by concepts like entanglement entropy from quantum information and numerical tools for analyzing (1+1)D quantum many-body systems, such as the density-matrix renormalization group (DMRG)~\cite{White:1992} and tensor-network ansatz~\cite{Vidal:2003,Vidal:2004}.
We refer to this modern reformulation as the tensor-network renormalization group (TNRG).
In TNRG, the partition function is encoded in a local tensor capturing the Boltzmann weight, as well as how these local tenors connect with each other to form a tensor network.
%{\color{orange}
Take the square-lattice Ising model with nearest-neighbor (NN) interaction as an example.
One tensor-network representation of its partition function is obtained by grouping the NN interactions between spins according to which shaded plaquette an interaction is located in, 
\begin{align}
    \label{eq:spin2TN1}
    &\includegraphics[scale=0.06, valign=c]{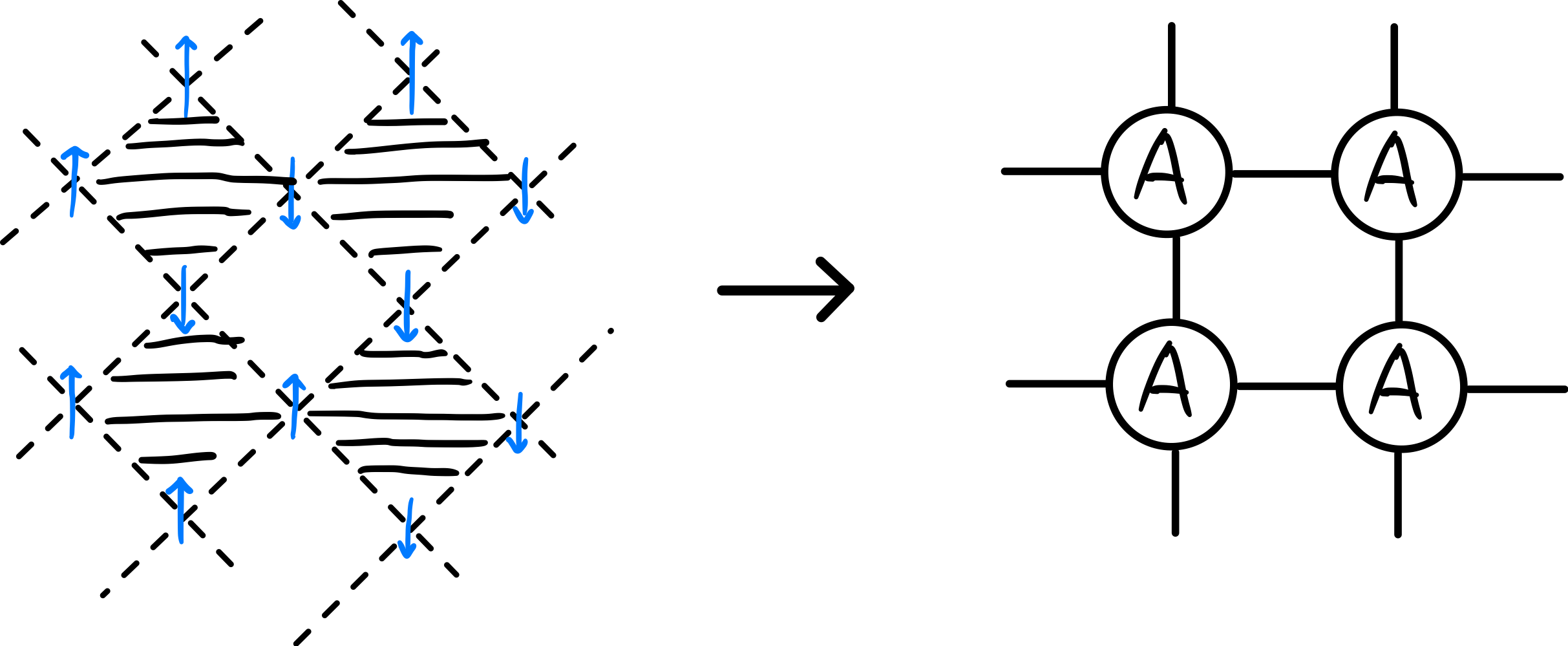}\quad,
    \text{with} \nonumber\\
    &\includegraphics[scale=0.04, valign=c]{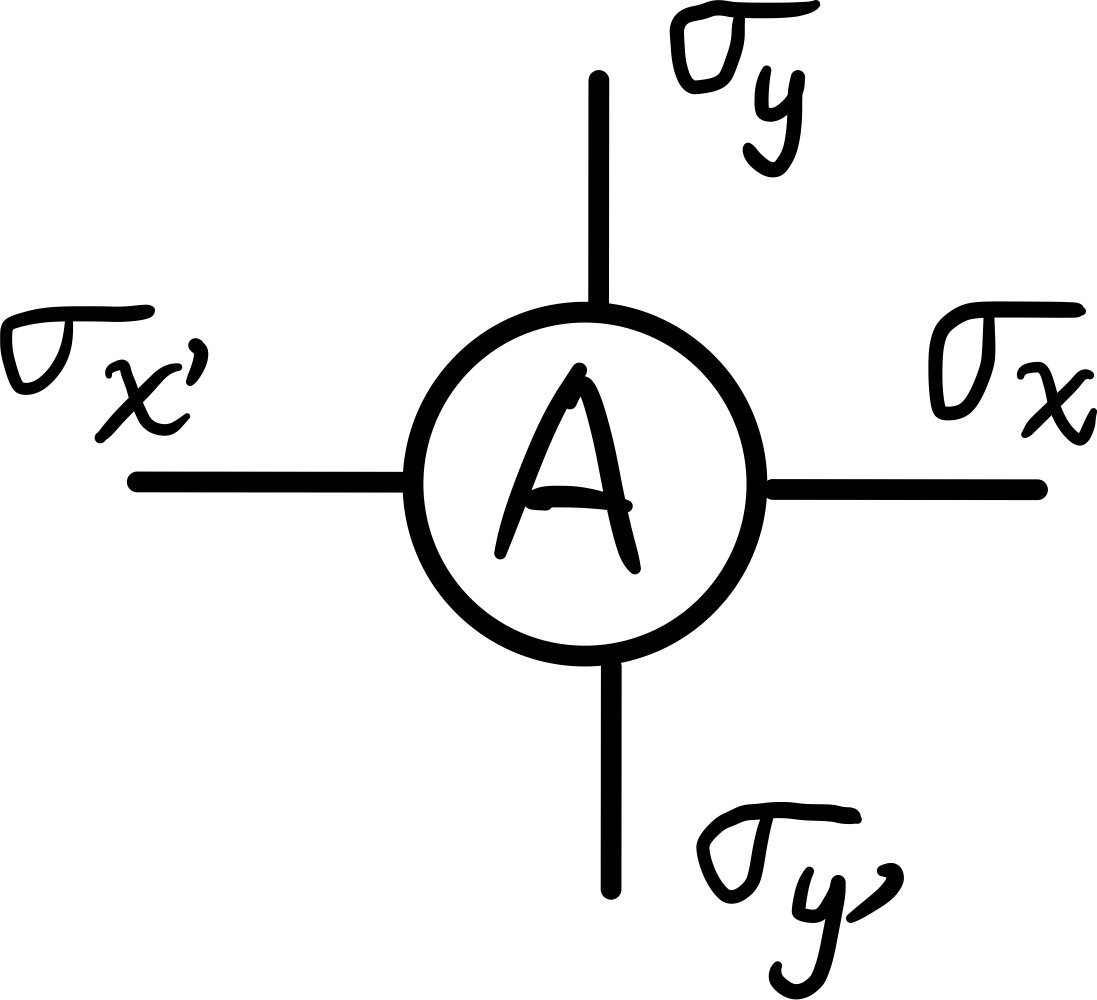}
    \equiv A_{\sigma_x \sigma_{x'} \sigma_{y} \sigma_{y'}}
    =
    e^{
        \beta (
        \sigma_x \sigma_y + \sigma_{y} \sigma_{x'} +
        \sigma_{x'}\sigma_{y'} + \sigma_{y'}\sigma_{x}
        )
},
\end{align}
where the blue up and down arrows denote spin variables $\sigma_i = \pm 1$ and $\beta$ is the inverse temperature.
In a tensor-network representation, if a tensor leg is shared by two tensors, it is summed over, or is contracted. 
This is similar to Einstein’s summation convention in tensor analysis. 
For example, the conventional mathematical expression for the following tensor-network contraction is
\begin{align}
    \label{eq:tenNotatEx}
    \includegraphics[scale=0.05, valign=c]{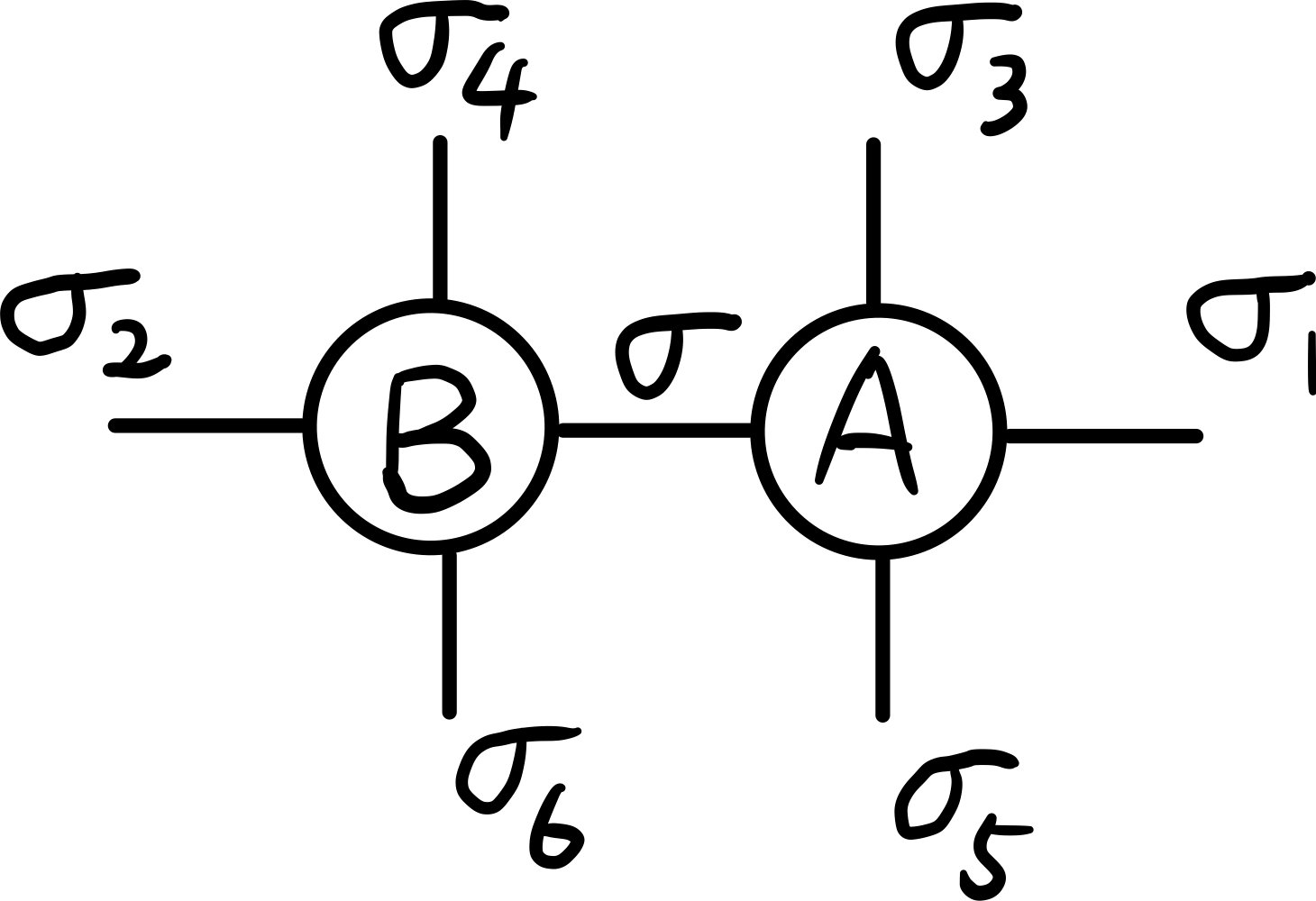}
    =
    \sum_{\sigma} A_{\sigma_1 \sigma \sigma_3 \sigma_5} B_{\sigma \sigma_2 \sigma_4 \sigma_6}.
\end{align}
The meaning of similar diagrammatic equations in the following should be clear.
If we impose the periodic boundary condition in both directions in the tensor network, the partition function of the square lattice Ising model is represented by a full contraction of the tensor network,
\begin{align}
    \label{eq:tn2Z}
    Z=
    \includegraphics[scale=0.10, valign=c]{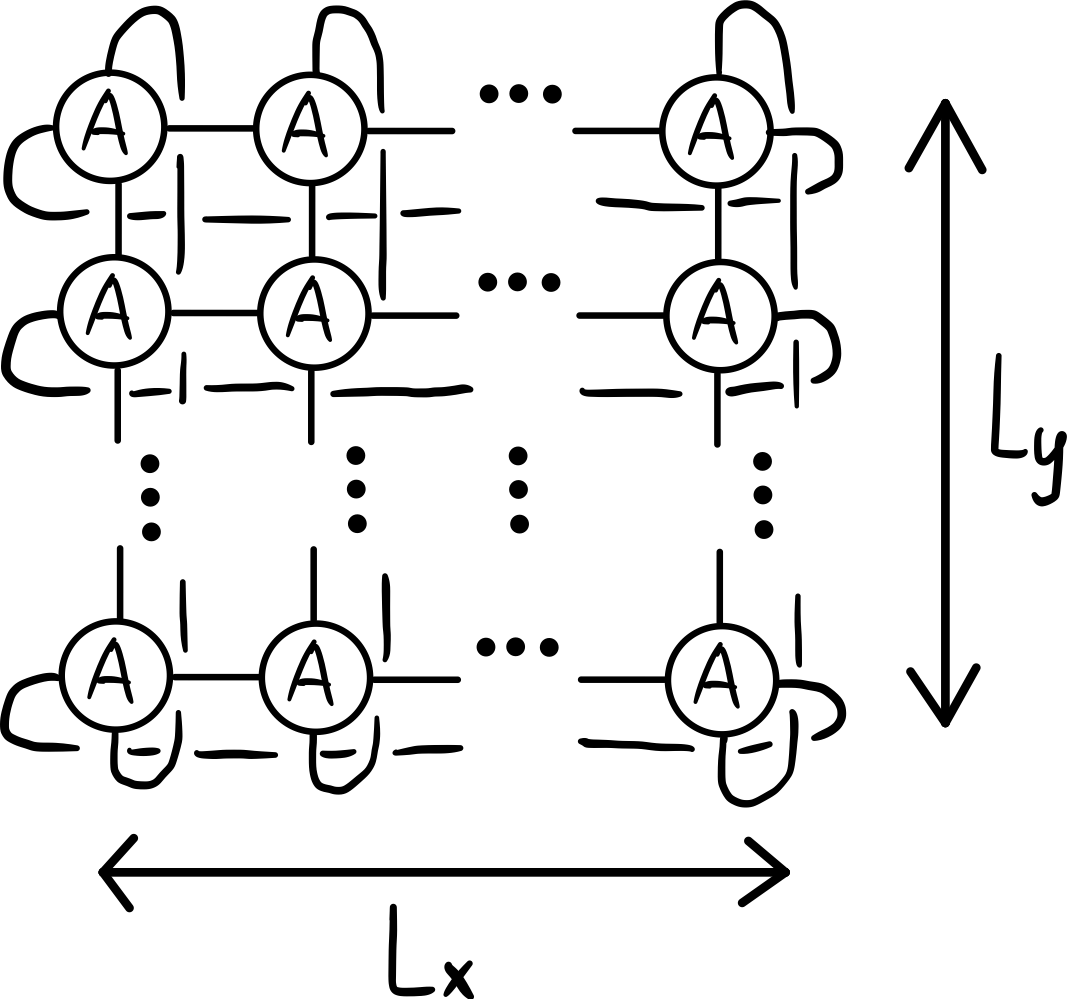}
    \equiv
    Z(A, L_x, L_y).
\end{align}
The number of tensors in $Z(A, L_x, L_y)$ is $N_{\text{tensor}} = L_x L_y$, while the number of spins is $N_{\text{spin}} = 2N_{\text{tensor}} = 2 L_x L_y$.
One advantage of a tensor-network representation is that it is possible to define a nature metric to quantify the RG approximation errors.
A detailed review can be found in Ref.~\cite{Evenbly:TNRalgo, Levin:Nave:2007, Lyu:2023}.
%}

%{\color{orange}
In tensor-network representation, Kadanoff's block idea becomes a block-tensor transformation.
For example, in a square-lattice tensor network in Eq.~\eqref{eq:tn2Z}, an exact block-tensor map with rescaling factor $b=2$ is obtained by contracting a $2 \times 2$ block of tensors,
\begin{align}
    \label{eq:bkten}
    \includegraphics[scale=0.08, valign=c]{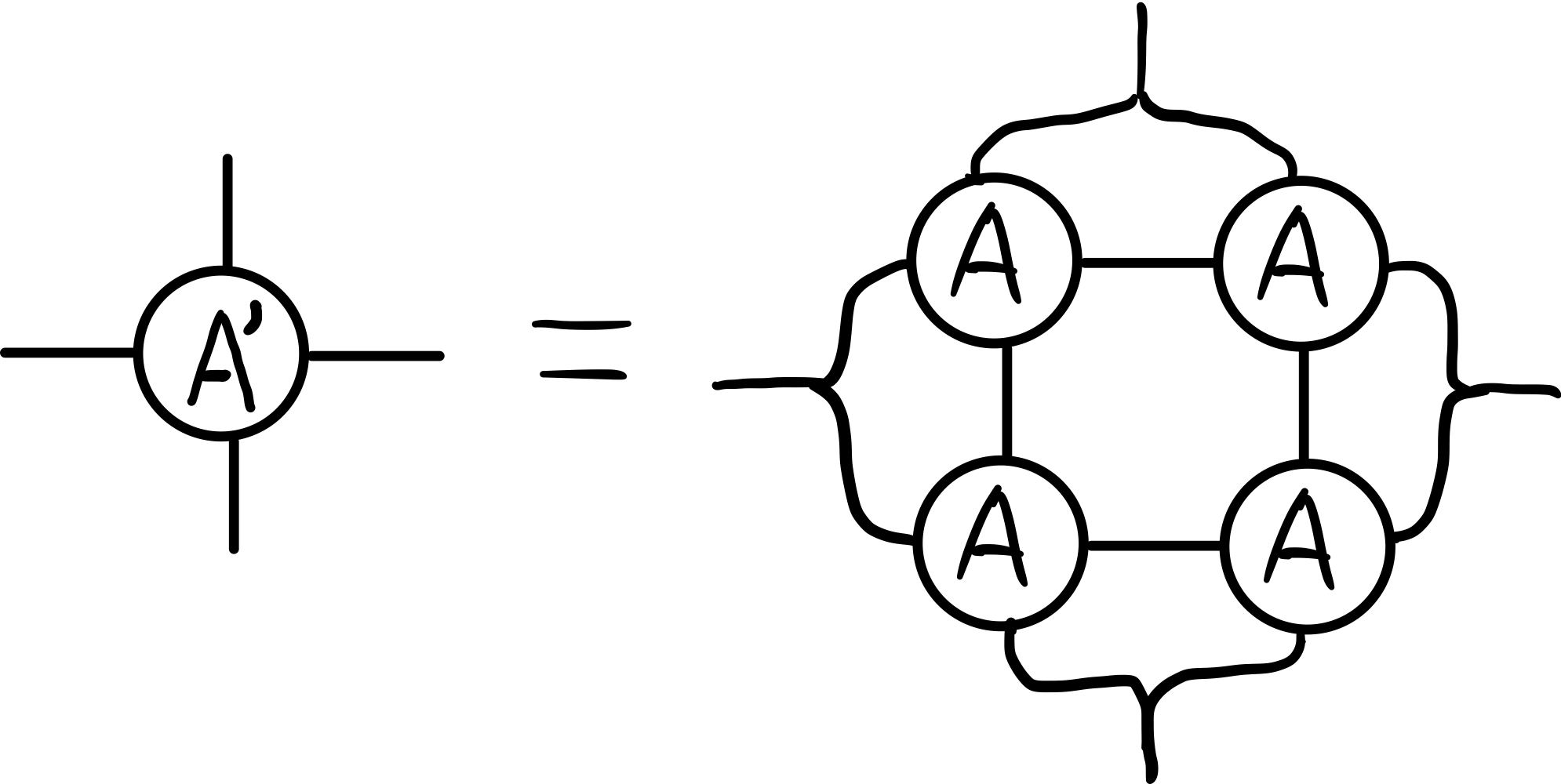}\quad,
\end{align}
where the merging of two legs means grouping two indices and relabeling them, and $A'$ is the coarse-grained tensor.
The partition function is invariant under this exact block-tensor map $Z(A, L_x, L_y) = Z(A', L_x/2, L_y/2)$.
Notice that the dimensionality of the vector space corresponding to a tensor leg grows from $2$ to $4=2^2$.
In numerical calculations, a truncation scheme is necessary to prevent this growth of dimensionality.
%}
Each index (or leg) has a finite dimensionality $\chi \in \mathbb{Z}$ called the bond dimension, which corresponds to the number of coupling constants retained during the RG\@.
There are many practical truncation schemes~\cite{Levin:Nave:2007,Xie:2012,Adachi:2020,Kadoh:2019,Adachi:2022} for this block-tensor transformation, all of whose basic ideas dates back to the DMRG\@.
In 2D, these methods inherit the high accuracy of the DMRG for (1+1)D quantum systems, according to the quantum/classical map~\cite{Kogut:1979,Shankar:2017}; their success is assured for non-critical systems due to the saturation of entanglement entropy~\cite{Levin:Nave:2007,Lyu:2023}.
Similar to the DMRG, what measures the RG approximation error in the TNRG is how fast the eigenvalues of some density matrix decay.
%{\color{orange}
    By repeating a tensor RG map $A \mapsto A'$ like Eq.~\eqref{eq:bkten}, a RG flow in the space of 4-leg tensor is generated,
    \begin{align}
        \label{eq:tenRGflow}
        A^{(0)} \mapsto A^{(1)} \mapsto \ldots
        \mapsto A^{(n)} \mapsto \ldots,
    \end{align}
    where $A^{(0)}$ denotes the initial tensor like the one in Eq.~\eqref{eq:spin2TN1}, while $A^{(n)}$ is the coarse-grained tensor after $n$ RG steps.
    After several RG steps, the size $L_x, L_y$ can be reduced to $O(1)$ and the partition function can be evaluated.
%}
The relative error of the estimated free energy of the square-lattice Ising model can easily go down to $\sim 10^{-9}$ using a personal computer~\cite{Evenbly:trgcode}.

%{\color{orange}
    The reason why entanglement entropy, which is defined for quantum systems, is useful in the context of a real space RG for classical statistical systems is due to a mixture of quantum and classical perspectives adopted by the TNRG~\cite{Levin:Nave:2007}.
    According to the physical picture of the tensor network representation of a partition function shown in Eq.~\eqref{eq:spin2TN1}, under the block-tensor map in Eq.~\eqref{eq:bkten}, four squares of block spins are tiled and spins in the bulk are summed over, leaving the boundary spin not summed.
    Therefore, a coarse-grained tensor $A^{(n)}$ can be seen as the wave function of a quantum system living on the boundary of a square of linear size $2^n$.
    Using this interpretation, the block-tensor RG can be regarded as an imaginary-time evolution along the radial direction.
    When the RG step $n$ is large, the wave function gets projected onto the ground state of the quantum system.
    This mixture of quantum and classical perspectives can help understand the properties of block-tensor maps using area laws of entanglement entropy~\cite{Lyu:2023}.
%}

% Entanglement filtering in 3D
\section{Entanglement filtering in 3D}
The justification for the success of the 2D block-tensor schemes no longer works in 3D due to the growth of entanglement entropy~\cite{Lyu:2023}.
This is not because of the much higher computational complexity in 3D, but because of a qualitative difference between 2D and 3D in their entanglement structure.
In our numerical experiment using the go-to block-tensor map in 3D, the higher-order tensor renormalization group (HOTRG)~\cite{Xie:2012}, we observe that for bond dimension $\chi \leq 22$, the RG errors grow rapidly to more than $10\%$ just after one RG step, and then keep growing to more than $30\%$ near the critical fixed point of the cubic-lattice Ising model.
Moreover, the RG errors near the critical region \emph{grow slowly} (not decrease!) with the bond dimension, which makes unreliable the block-tensor map in 3D.

\par
A way out for this growth of entanglement entropy is to enhance the simple block-tensor transformation with an entanglement filtering (EF) process~\cite{Gu:Wen:2009,Vidal:2007,Evenbly:Vidal:2015}.
If designed carefully, an EF scheme can significantly reduce the entanglement entropy in the coarse-grained description and hence reduce the RG errors of the block-tensor transformation~\cite{Vidal:2007,Evenbly:Vidal:2015}.
There are many EF schemes\cite{Vidal:2007,Gu:Wen:2009,Evenbly:Vidal:2015,Yang:Gu:Wen:2015,Bal:2017,Hauru:Delcamp:Mizera:2018,Evenbly:2018,Ying:2017,Harada:2018} in 2D but a workable 3D scheme is still missing.

\par
\emph{Two ideas are essential for a 3D EF scheme: graph independence and imposing the necessary symmetries of the microscopic model.}
If an EF process does not alter the graph formed by the tensor network, it is called graph independent~\cite{Hauru:Delcamp:Mizera:2018}.
Graph independence makes an entanglement filtering a standalone procedure, which can be incorporated into any block-tensor schemes.
This versatility is essential for designing a 3D EF-enhanced TNRG\@.

\par
To obtain a critical fixed-point tensor, it is helpful to exploit necessary symmetries of the microscopic model in the TNRG.
Without symmetries being imposed, even if we start from symmetric initial tensors, numerical errors associated with machine precision, or even worse, with the artifacts of a chosen scheme, introduce all possible perturbations away from the critical model located on the critical surface.
Since a few RG steps are needed to drive this critical model to the corresponding critical fixed point, those perturbations along the RG relevant directions inevitably get amplified.
This entails difficulty in obtaining a critical fixed-point tensor.
People have developed mature framework for imposing global on-site symmetry~\cite{Singh:2010}, like the spin-flip $\mathbb{Z}_2$ symmetry of the Ising model~\cite{Singh:2011}.
For the 2D Ising model, it is enough to impose the spin-flip $\mathbb{Z}_2$ symmetry because the only RG relevant direction in the spin-flip even sector is the energy density field $\epsilon$, which corresponds to the fine-tuning of the temperature to its critical value $T_c$.
For the 3D Ising model, however, there are more RG relevant directions.
Besides the energy density field $\epsilon$, whose scaling dimension is about 1.41, its three first descendants, $\partial_x \epsilon, \partial_y \epsilon, \partial_z \epsilon$, have scaling dimension about 2.41, smaller than the spatial dimension 3~\cite{Rychkov:2017}.
%{\color{orange}
    Although these derivative fields can have arbitrary eigenvalues in the Jacobian of the RG map~\cite{Ebel:2024LDO}, which are not related to the CFT values, the fact that they could be these values still cause some worries in numerical calculations.
%}

\par
In our numerical experiment, adding a naive 3D generalization of the scheme in Ref.~\cite{Hauru:Delcamp:Mizera:2018} to the HOTRG makes it harder to obtain a critical fixed-point tensor.
We conjecture that it is because this naive generalization breaks the lattice reflection and rotation symmetry explicitly; this numerical artifact could introduce perturbations in directions of $\partial_x \epsilon,\partial_y \epsilon,\partial_z \epsilon$.
To eliminate these directions in the RG map, we develop techniques for imposing lattice reflection symmetry in TNRG\@.
The unique feature of our proposed EF scheme is that it is both graph independent and with the lattice reflection symmetry imposed.
\emph{The gist of the proposed EF is captured in the following approximation in 2D},
\begin{align}
    \label{eq:EF2D-approx}
    \includegraphics[width=0.8\columnwidth, valign=c]{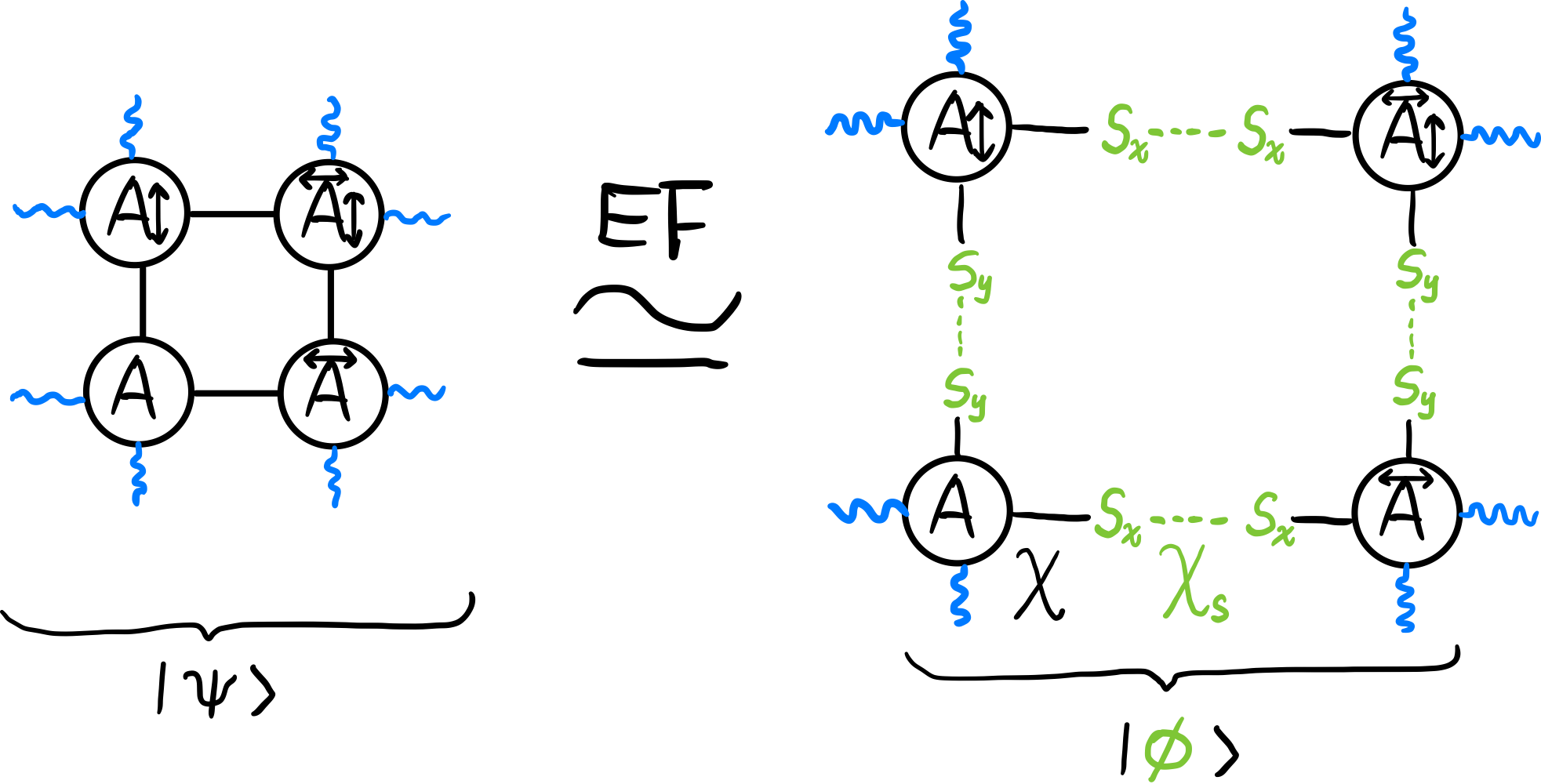},
\end{align}
where the two filtering matrices $s_x, s_y$ squeeze the bond dimension from $\chi$ to $\chi_s < \chi$, while we want the plaquette after this squeezing remains as close as the original one.
The double arrow along the horizontal or the vertical axis means the two legs along that direction are transposed.
For example, $\includegraphics[scale=0.02, valign=c, raise=0em]{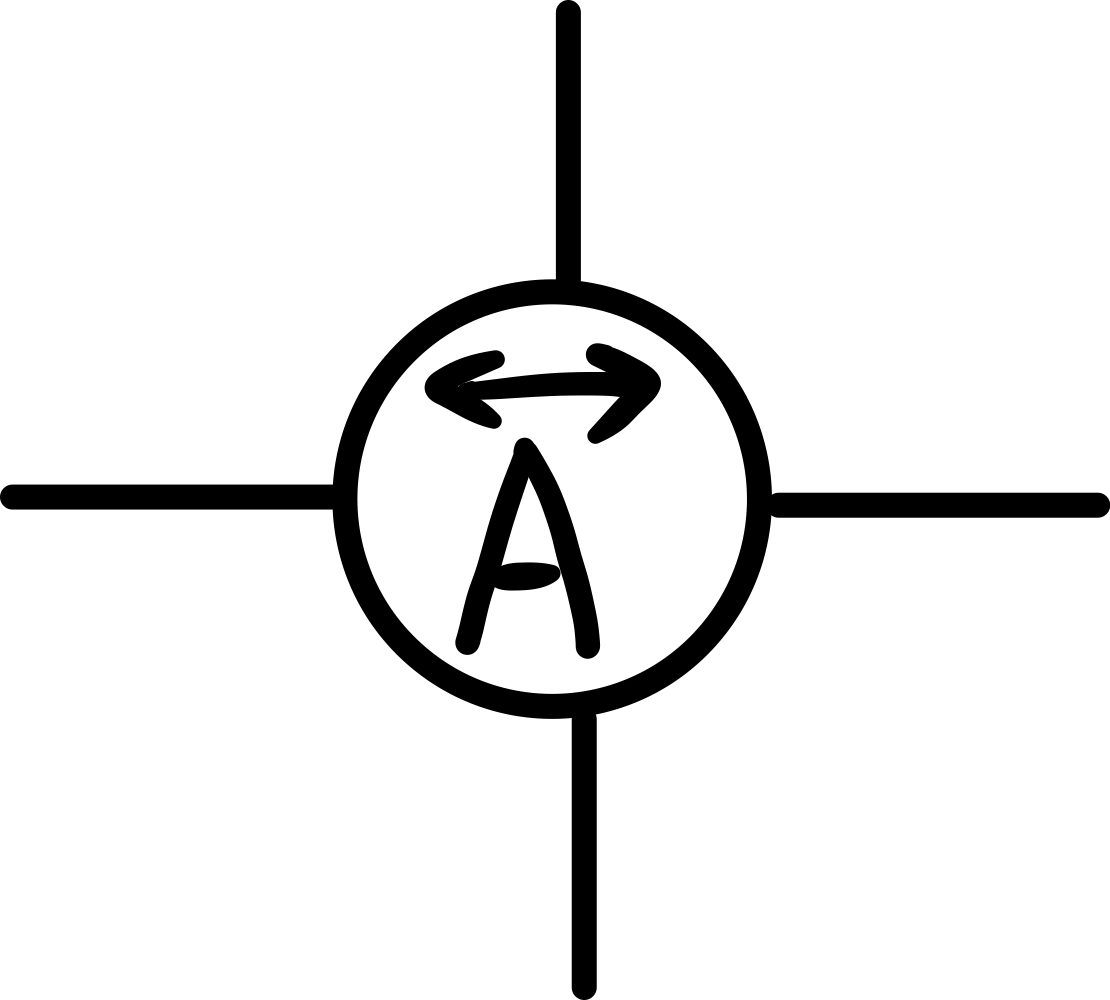} \equiv \includegraphics[scale=0.02, valign=c, raise=0em]{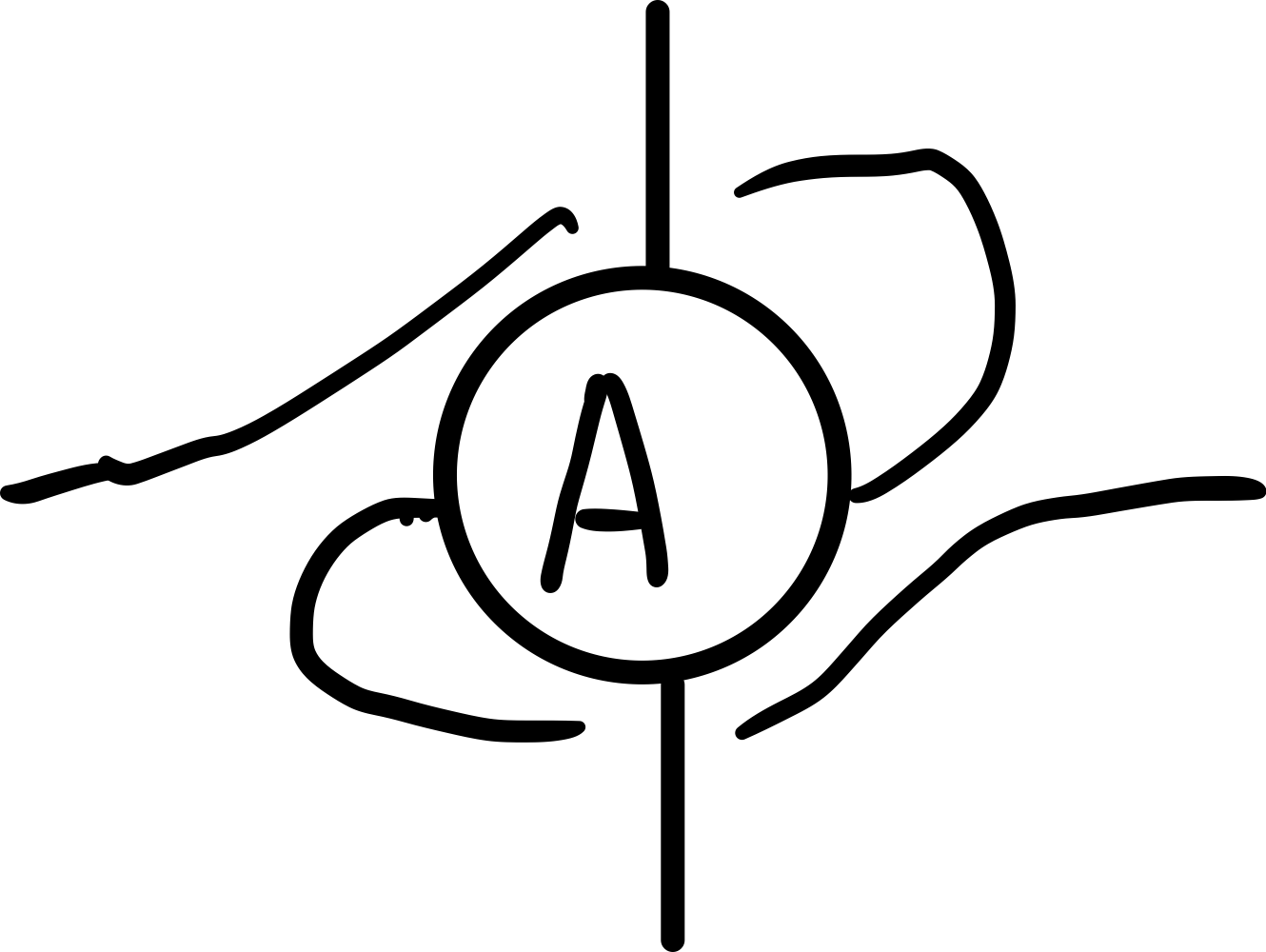}$.
The reason why this transposition trick can exploit lattice reflection symmetry of the underlying model will be explained in a coming paper.
%{\color{orange}
    After the four black solid legs of the plaquette on the left-hand side of Eq.~\eqref{eq:EF2D-approx} are contracted, the tensor network can be seen as a vector $\ket{\psi}$ living in a vector space corresponding to the 8 wavy legs~\footnote{
    %{\color{orange}
        When a tensor network is used as a ground state ansatz, like a matrix product state (MPS) or a projected entangled pair state (PEPS), the blue wavy legs on the left-hand side of Eq.~\eqref{eq:EF2D-approx} are called physical legs, while the black solid legs virtual legs.
    %}
}.
Similarly, the tensor network on the right-hand side of Eq.~\eqref{eq:EF2D-approx} is treated as another vector $\ket{\phi}$ living in the same vector space.
%}
These two filtering matrices $s_x, s_y$ are initialized according to the technique in Ref.~\cite{Hauru:Delcamp:Mizera:2018} and further optimized using the method developed in Ref.~\cite{Evenbly:2018}, by maximizing the overlap between the filtered state $\ket{\phi}$ and the target state $\ket{\psi}$ in Eq.~\eqref{eq:EF2D-approx}.
The 3D generalization is in Eq.~\eqref{eq:EF3D}.

% Integrate EF into block-tensor map
\section{Integrating an EF into a block-tensor map}
A block-tensor transformation fails to simplify microscopic entanglement located near the intersection regions of groups of spins~\cite{Vidal:2007}.
In a 2D block-tensor map, these boundary entanglement transforms like
\begin{subequations}
    \label{eq:EntInBkten2D}
\begin{align}
    \label{eq:TN2spinpic}
    \includegraphics[width=0.4\columnwidth, valign=c]{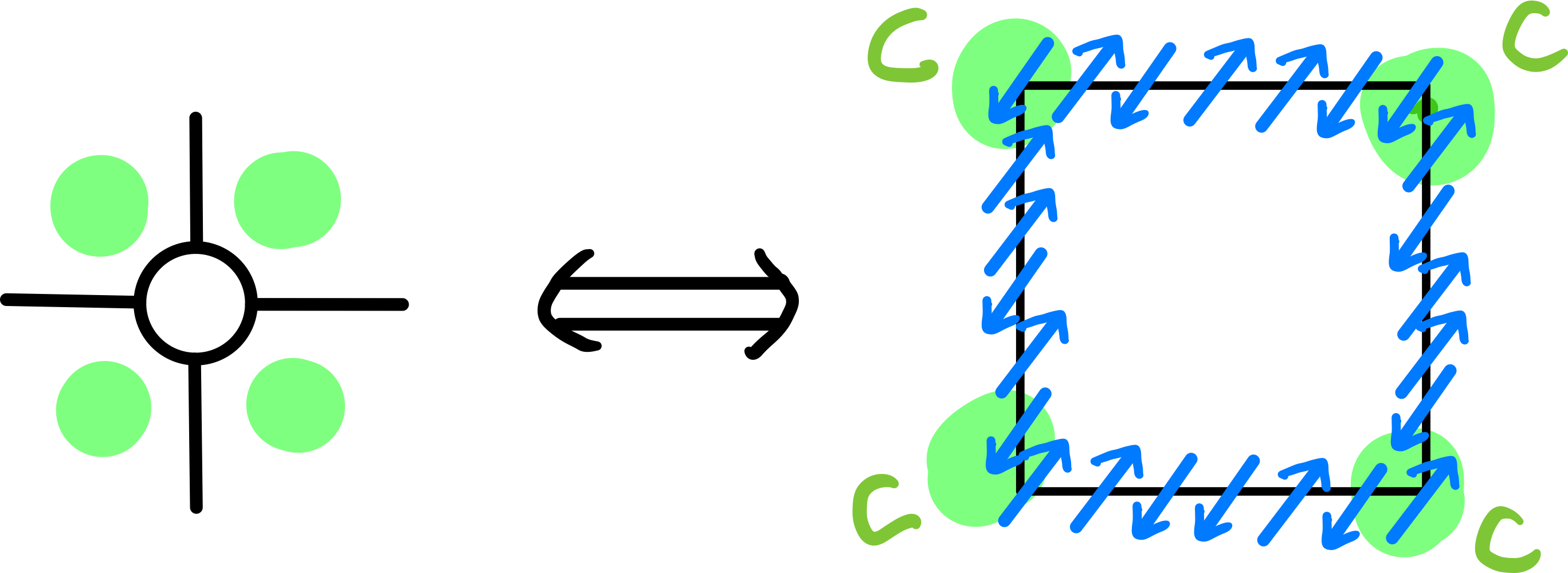}, \\
    \label{eq:block2spinpic}
    \includegraphics[width=0.8\columnwidth, valign=c]{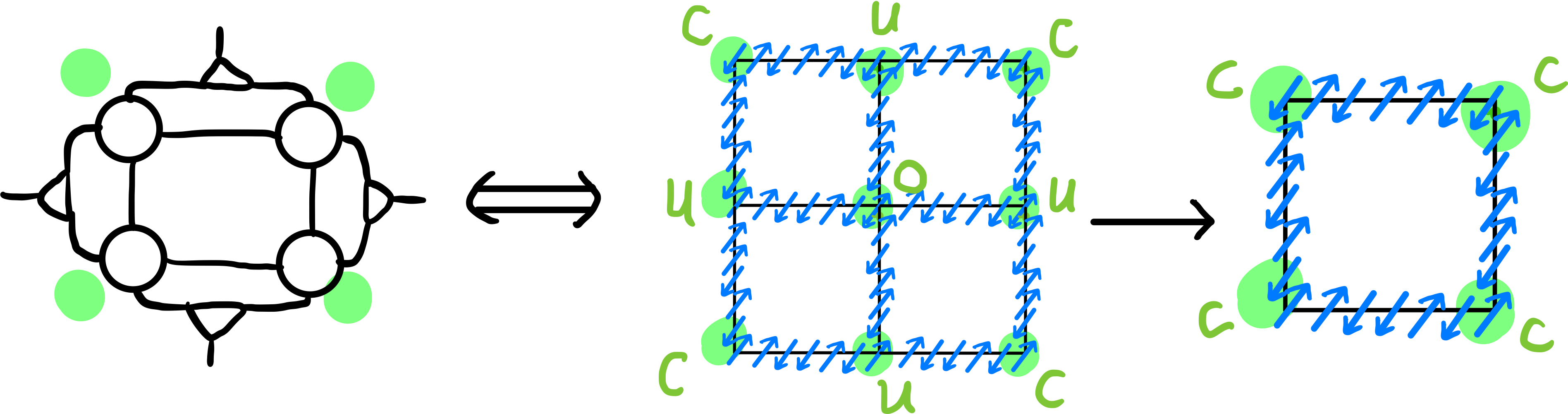}.
\end{align}
\end{subequations}
These two equations are intuitive in nature, whose more precise form can be understood using corner-double-line tensors~\cite{Levin:Nave:2007,Gu:Wen:2009}.
The correspondence in Eq.~\eqref{eq:TN2spinpic} demonstrates the relationship between the 4-leg tensor and the original spins, where the blue arrows denotes the original spin degrees of freedom and the shaded dots $c$ marks the location of boundary entanglement.
Under the block-tensor transformation in Eq.~\eqref{eq:block2spinpic}, the entanglement among the spins in the region around the center of the block denoted by $o$ is renormalized to a single number after the spins on the inner edges are summed over. 
The entanglement around the center of the edge denoted by $u$ is among spins on the same edge of a larger block, and thus will be renormalized after a isometric transformation in the block-tensor map~\cite{Evenbly:Vidal:2015,Lyu:2021}. 
The entanglement located around four corners denoted by $c$ behaves differently and fails to be eliminated under the block-tensor map. 
Equation~\eqref{eq:EntInBkten2D} is the tensor-network incarnation of the entanglement-entropy area law~\cite{Lyu:2023}.

\begin{figure}[tb]
    \includegraphics[width=1.00\columnwidth,valign=c]{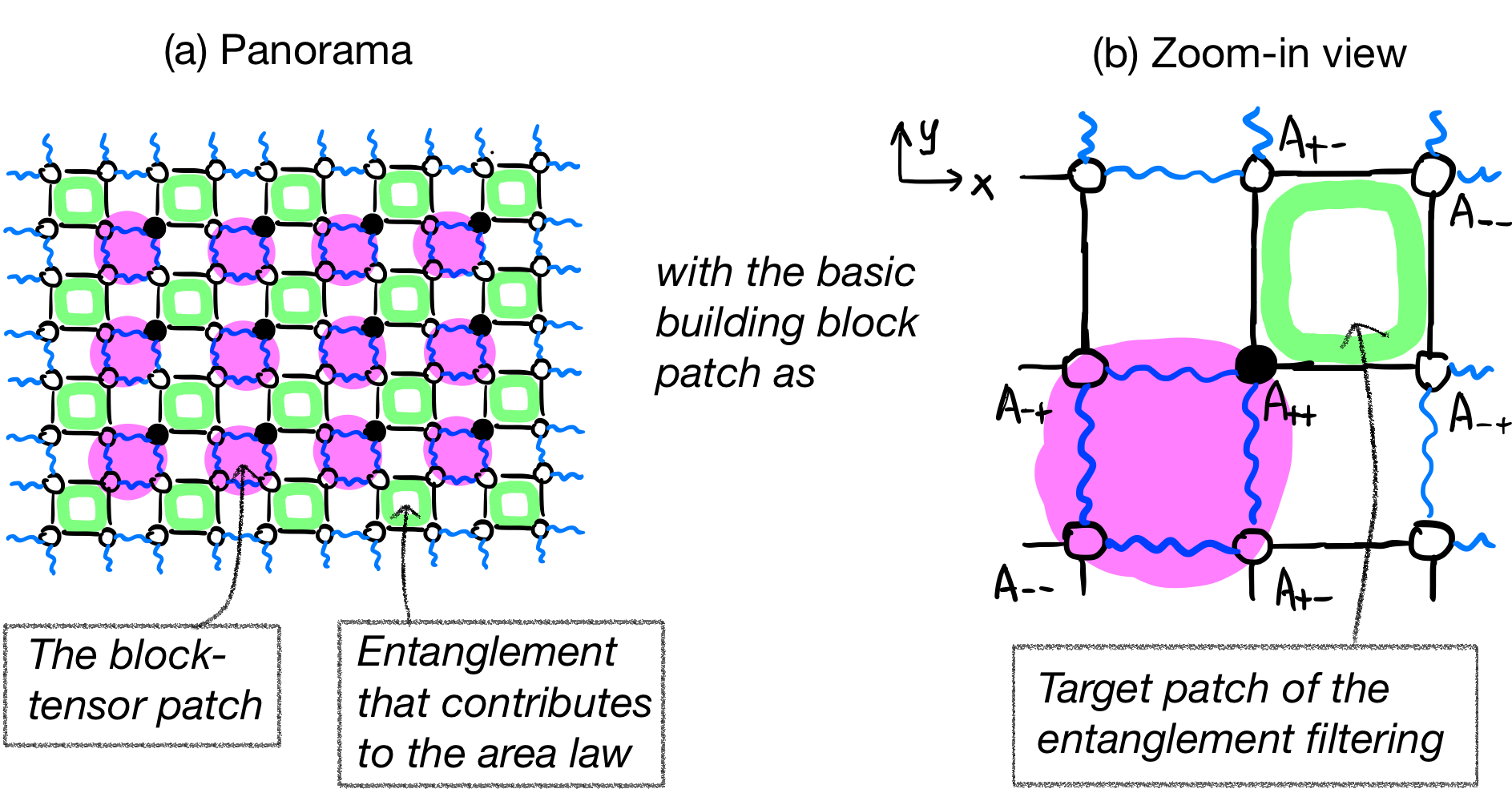}
    \caption{\label{fig:panorama2D}
        Location of the leftover entanglement for a 2D block-tensor transformation.
}
\end{figure}

\begin{figure}[tb]
    \includegraphics[width=0.95\columnwidth,
    % left,
    valign=c]{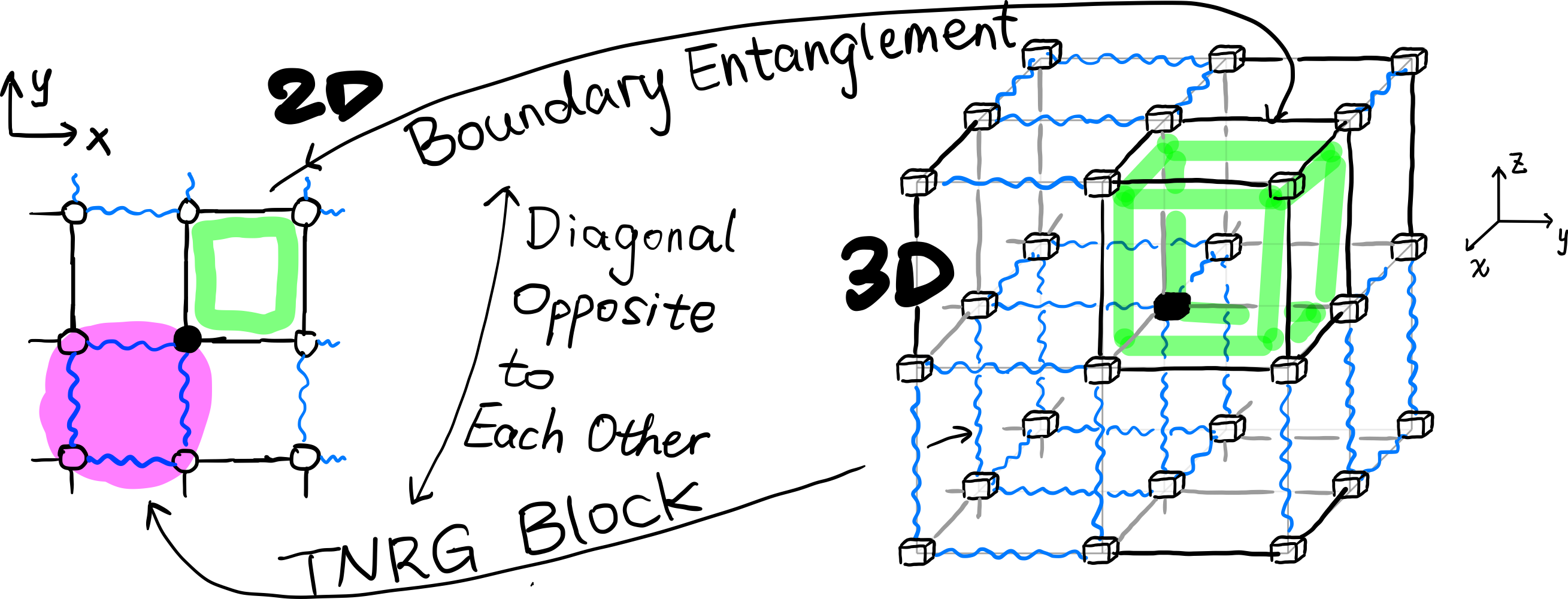}
    \caption{\label{fig:panorama3D}
        The general principle for integrating entanglement filtering into a block-tensor map in any spatial dimensionality, demonstrated in the 2D and the 3D.
}
\end{figure}

\par
Therefore, the EF should target patches of tensor network where the short-range entanglement fails to be eliminated under the block-tensor map (see~\autoref{fig:panorama2D}).
Let us write down a principle of how to integrate the EF into a block-tensor map in 2D (see the 2D diagram in~\autoref{fig:panorama3D}), in such a way that the 3D generalization is straightforward:
\par
Step 1: Choose an \emph{anchor point} in the $2 \times 2$ block tensor patch; we indicate it using a black dot at the center;
put the main tensor at the anchor point.
\par
Step 2: Draw the two legs of the main tensor that points to the positive $x$ and $y$ directions as black lines and call them “outer legs”;
these two legs span the plaquette that is the target of the entanglement filtering (see~\autoref{fig:panorama2D}).
\par
Step 3: Draw the two legs of the main tensor that points to the negative directions as blue wavy lines and call them “inner legs”; 
these two legs span the plaquette that is under the block tensor transformation (see~\autoref{fig:panorama2D}).

\par
In summary, \emph{the relationship between the block-tensor patch and the EF one is that they are diagonally opposite to each other with respect to the anchor point}. 
The generalization to 3D is straightforward (see~\autoref{fig:panorama3D}). 
Recently, an exact treatment of the 3D TNRG adopts a similar idea~\cite{Ebel:2024}.

% The proposed RG map
\section{The proposed RG transformation}
Apply the EF to the target patch shown in~\autoref{fig:panorama3D} and find good choices of the filtering matrices in three directions, $s_x, s_y, s_z$, according to the approximation (the anchor point is denoted as $(+++)$):
\begin{subequations}
    \label{eq:EF3D}
\begin{align}
    \label{eq:EF3Dapprox}
    \includegraphics[width=0.85\columnwidth, valign=c]{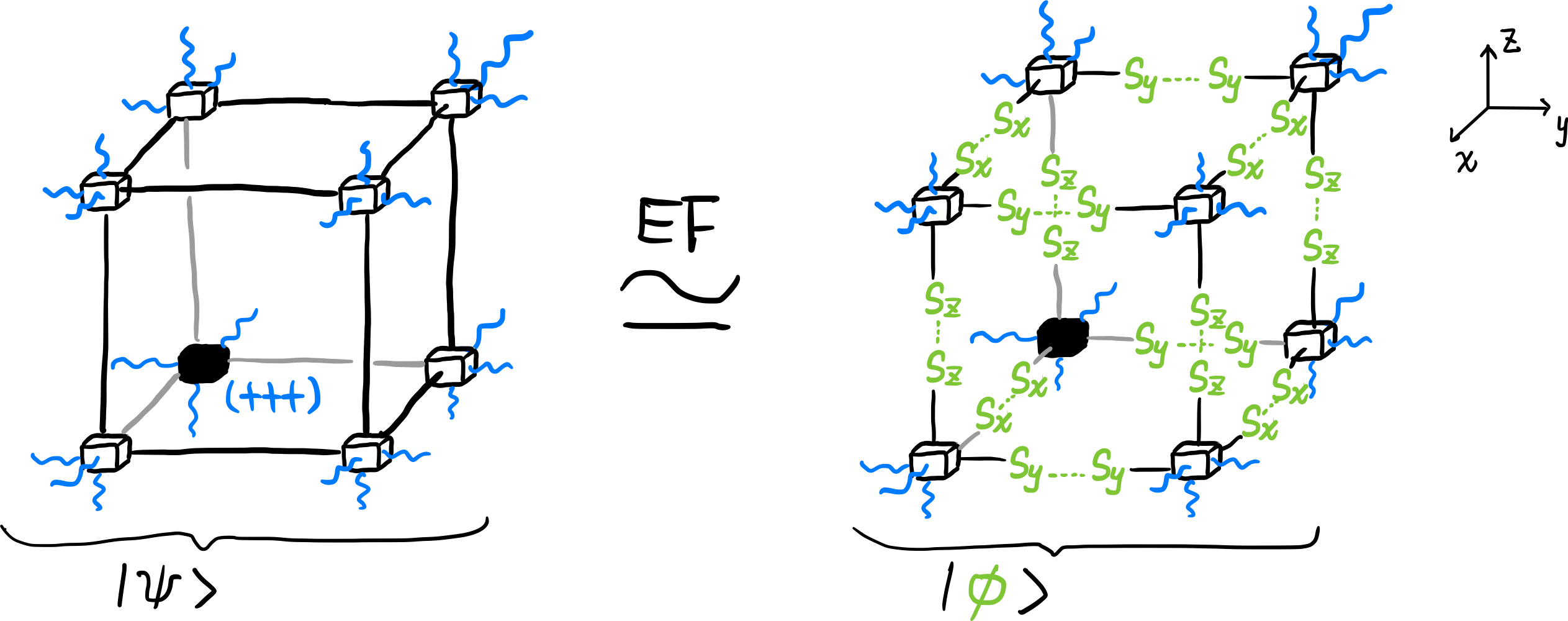}.
\end{align}
We call this a \emph{cube filtering} since the target patch is a $2 \times 2 \times 2$ cube.
%{\color{orange}
The lattice reflection symmetry is exploited in this step by a transposition trick similar to the one in Eq.~\eqref{eq:EF2D-approx}.
In Appendix~\ref{sec:app:reflsym}, we discuss more about lattice reflection symmetry and the transposition trick.
In Eq.~\eqref{eq:EF3Dapprox}, the 6-leg tensor $A$ is at the anchor point of the cube, while its different transpositions along various directions are at other seven points.
Take the left-hand side of Eq.~\eqref{eq:EF3Dapprox} as an example, the tensor $A$ and its transpositions are put into the cube according to
%}
\begin{align}
    \label{eq:transposeTrick}
    \includegraphics[width=0.85\columnwidth, valign=c]{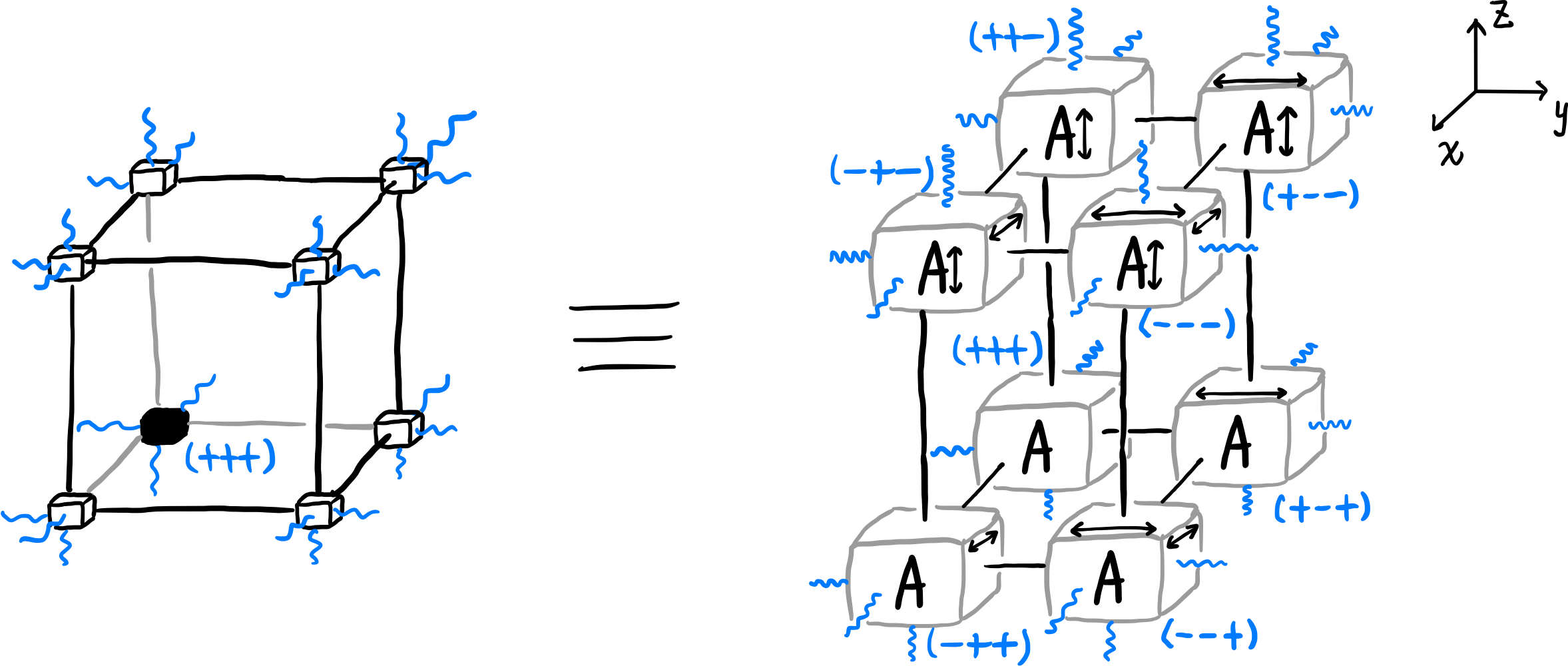}.
\end{align}
\end{subequations}
The diagrammatic notation in Eq.~\eqref{eq:EF3D} and the strategy for finding good filtering matrices have been outlined below Eq.~\eqref{eq:EF2D-approx}.
The computational costs of finding good $s_x, s_y, s_z$ are $O(\chi^{12})$.
Then, apply the filtering matrices on the three ``outer legs'' of the tensor at the anchor point, after which the TNRG block in~\autoref{fig:panorama3D} looks like
\begin{align}
    \label{eq:filtering}
    \includegraphics[width=0.9\columnwidth, valign=c]{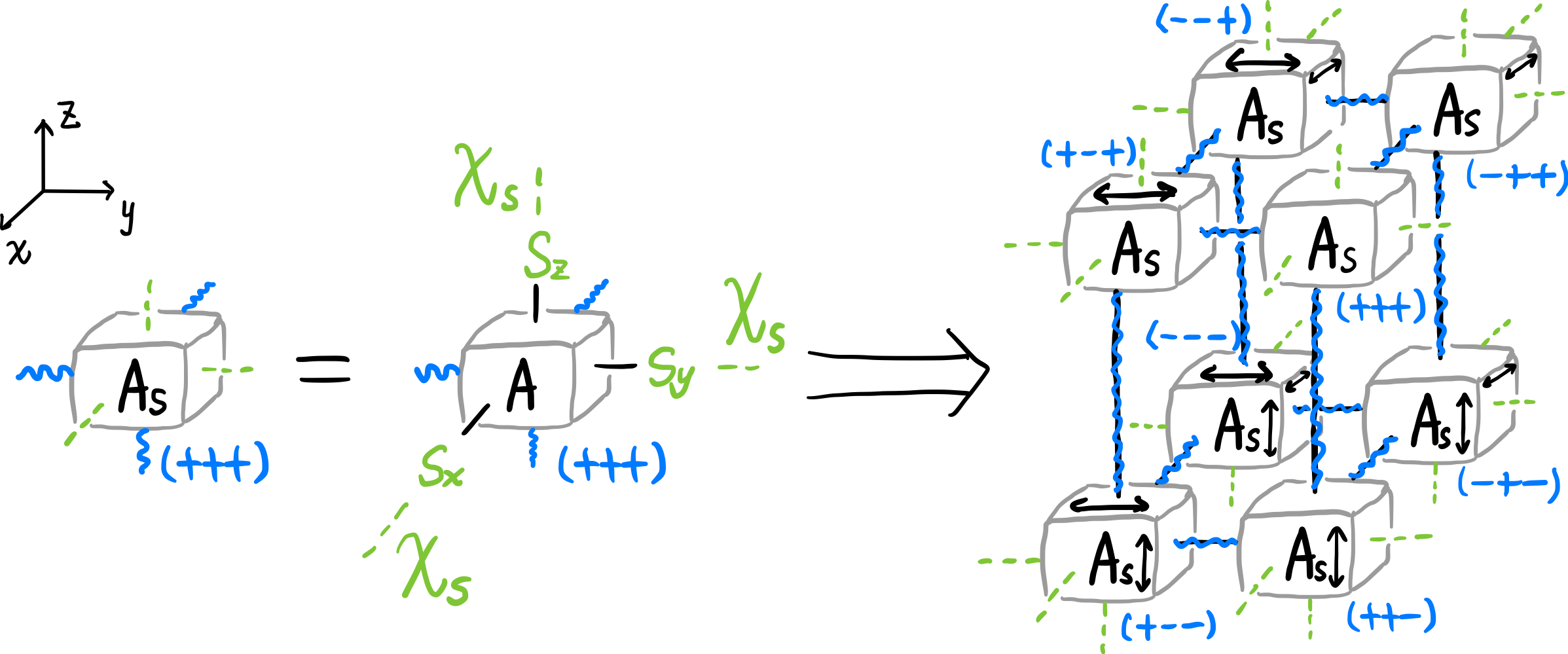}.
\end{align}
%{\color{orange}
This map $A \mapsto A_s$ is the first step of the proposed RG transformation.
%}
Notice that the position of the anchor point $(+++)$ in Eq.~\eqref{eq:filtering} is different from that in Eq.~\eqref{eq:transposeTrick}.
%{\color{orange}
    The reason is that the cube in Eq.~\eqref{eq:transposeTrick} is the entanglement patch in~\autoref{fig:panorama3D}, while the cube in Eq.~\eqref{eq:filtering} is the TNRG block.
%}
If the choice of filtering matrices are good, this step should filter out those 3D boundary entanglement structures discussed in Refs.~\cite{Hauru:Delcamp:Mizera:2018,Lyu:2023}.

\par
After the cube filtering, we apply an HOTRG-like block-tensor transformation to the TNRG block.
We adopt the HOTRG idea of coarse-graining one direction at a time and choose (arbitrarily) the order to be $z \rightarrow y \rightarrow x$.
%{\color{orange}
    The coarse-graining in each direction is accomplished by  inserting pairs of isometric tensors~\footnote{
        %{\color{orange}
        An isometric tensor $p_{ij\alpha}$ satisfies $\sum_{ij}p_{ij\alpha} p_{ij\beta} = \delta_{\alpha\beta}$.
        It fuses two tensor leg  $i,j$ into a single tensor leg $\alpha$.
        A pair of $p$ head-by-head contracting together, $\sum_{\alpha} p_{ij \alpha} p_{mn \alpha}$, forms a projection operator from index pair $(ij)$ to  $(mn)$, or vice versa.
        %}
    } into the tensor network (this technique is known as projective truncations~\cite{Evenbly:TNRalgo}).
%}
We demonstrate the $z$-direction collapse using the two tensors at the $(+++)$ and $(++-)$ positions in Eq.~\eqref{eq:filtering}.
%{\color{orange}
    Four types of isometric tensor $p_{mx}, p_{my}, p_{ix}, p_{iy}$, each coming in as two copies ``head-by-head'' contracting to each other, are inserted according to,
%}
\begin{subequations}
    \label{eq:hotrglikeBkten}
\begin{align}
    \label{eq:zblock}
    \includegraphics[width=0.80\columnwidth, valign=c]{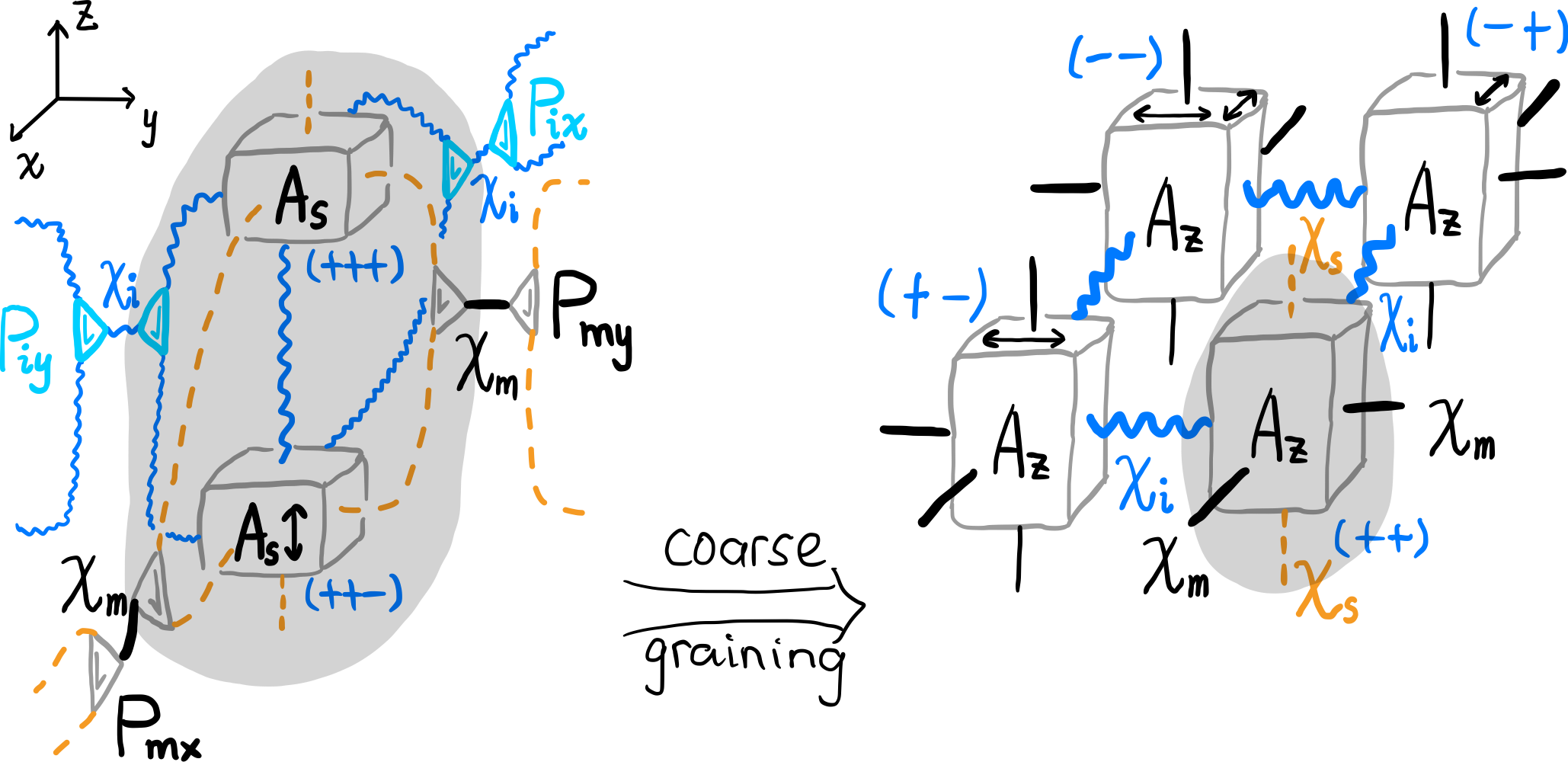} \nonumber\\
    \text{with} \quad
    \includegraphics[width=0.70\columnwidth, valign=c]{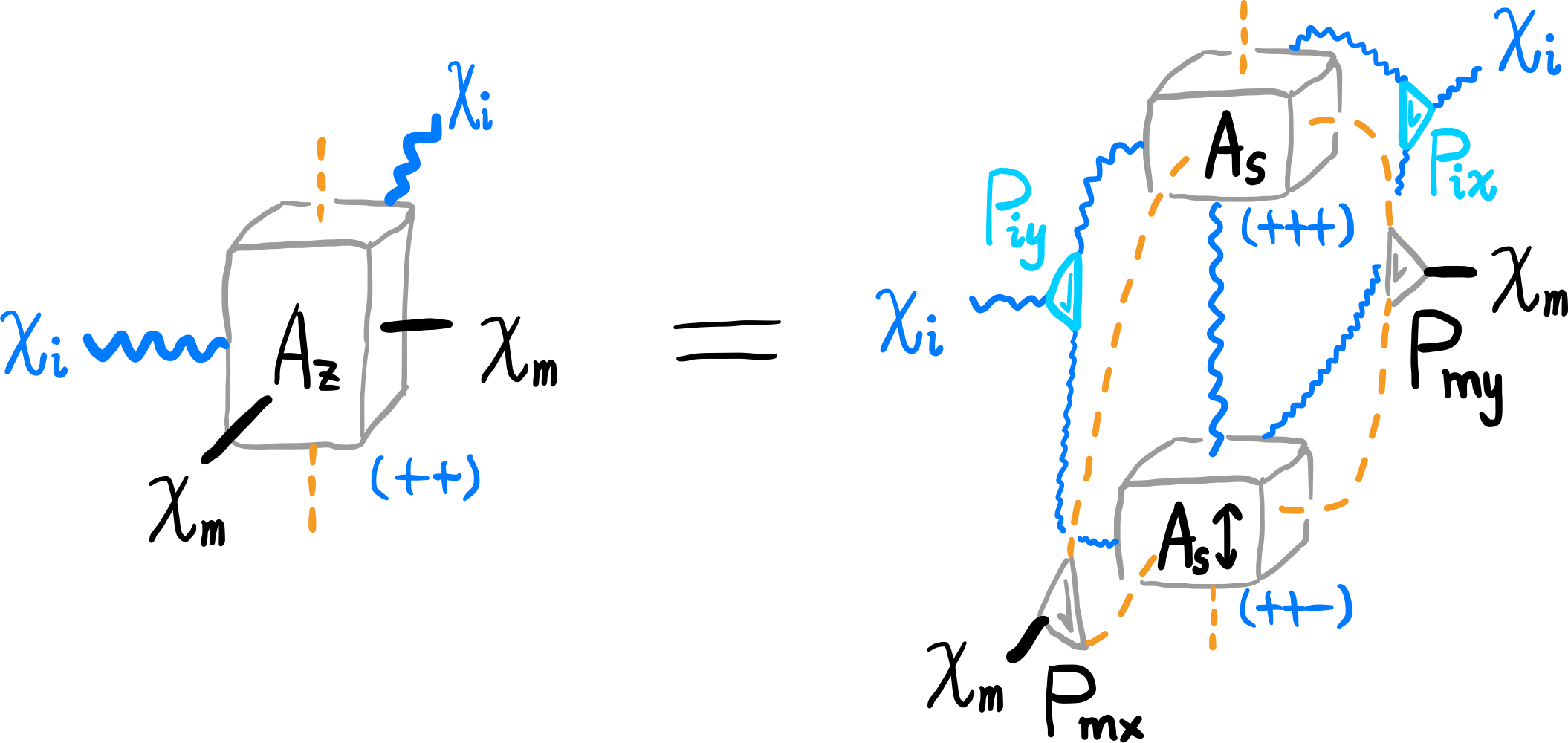}
\end{align}
%{\color{orange}
This map $A_s \mapsto A_z$ is the second step of the proposed RG transformation.
%}
The difference between the usual HOTRG and this one is that isometric tensors are different for inner and outer legs of the tensors in the block-tensor patch.
For the outer legs, the isometric tensors are $p_{mx}, p_{my}$, with the output bond dimension $\chi_{m}$; for the inner legs, they are $p_{ix}, p_{iy}$, with the output bond dimension $\chi_{i}$.
We use the projective truncations developed in Ref.~\cite{Evenbly:TNRalgo} to determine these isometric tensors.
For the $y$ direction, we focus on the tensors at the $(++)$ and $(+-)$ in Eq.~\eqref{eq:zblock}.
%{\color{orange}
Now, three types of isometric tensor $p_{mz}, p_{ox}, p_{iix}$ are needed,
%}
\begin{align}
    \label{eq:yblock}
    \includegraphics[width=0.80\columnwidth, valign=c]{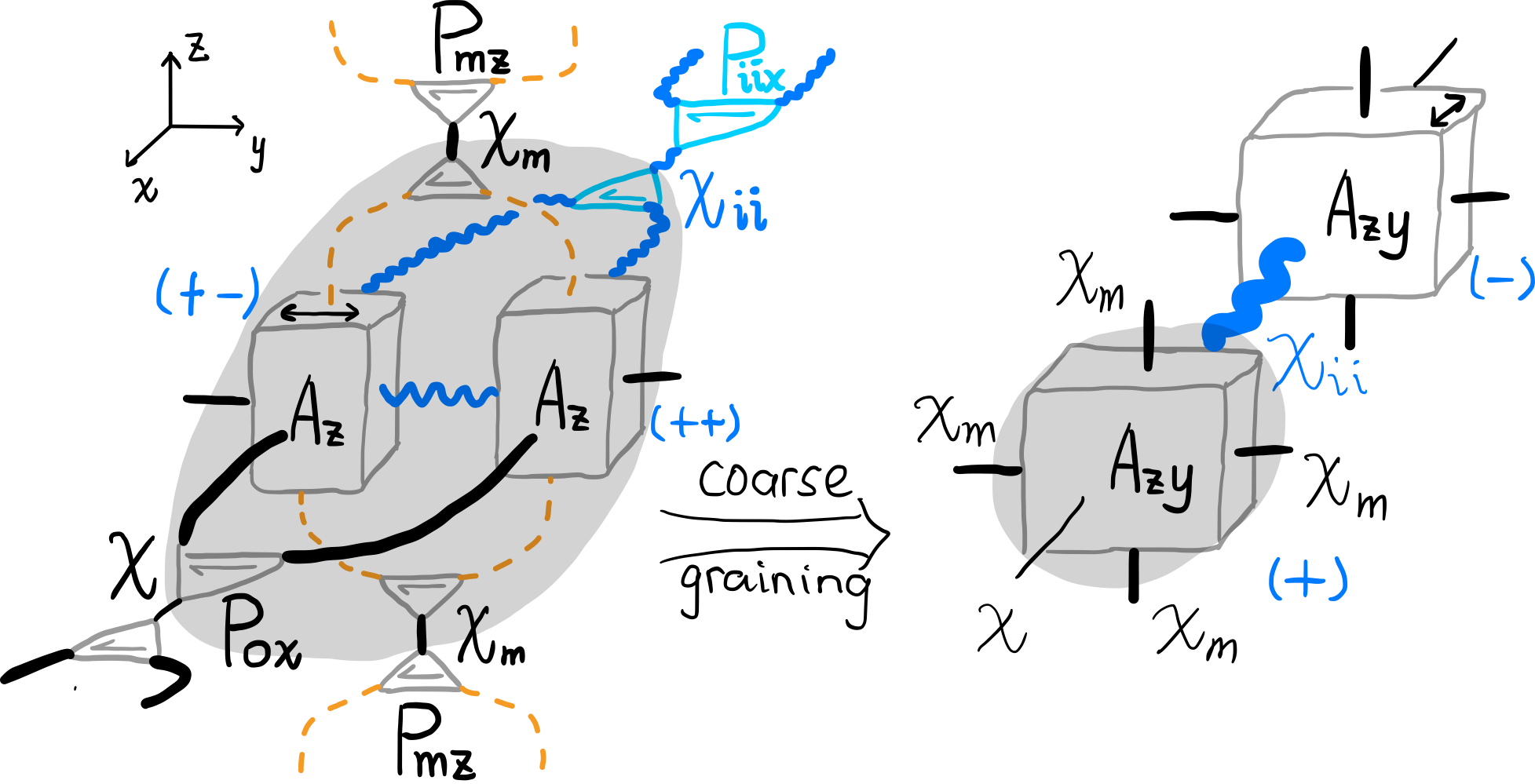},
\end{align}
%{\color{orange}
    with $A_{zy}$ obtained from the contraction of two copies of $A_z$ in a similar manner to the last line of Eq.~\eqref{eq:zblock}.
    This map $A_z \mapsto A_{zy}$ is the third step of the proposed RG transformation.
Finally, for the $x$ direction, we need two types of isometric tensors $p_{oy}, p_{oz}$,
%}
\begin{align}
    \label{eq:xblock}
    \includegraphics[width=0.80\columnwidth, valign=c]{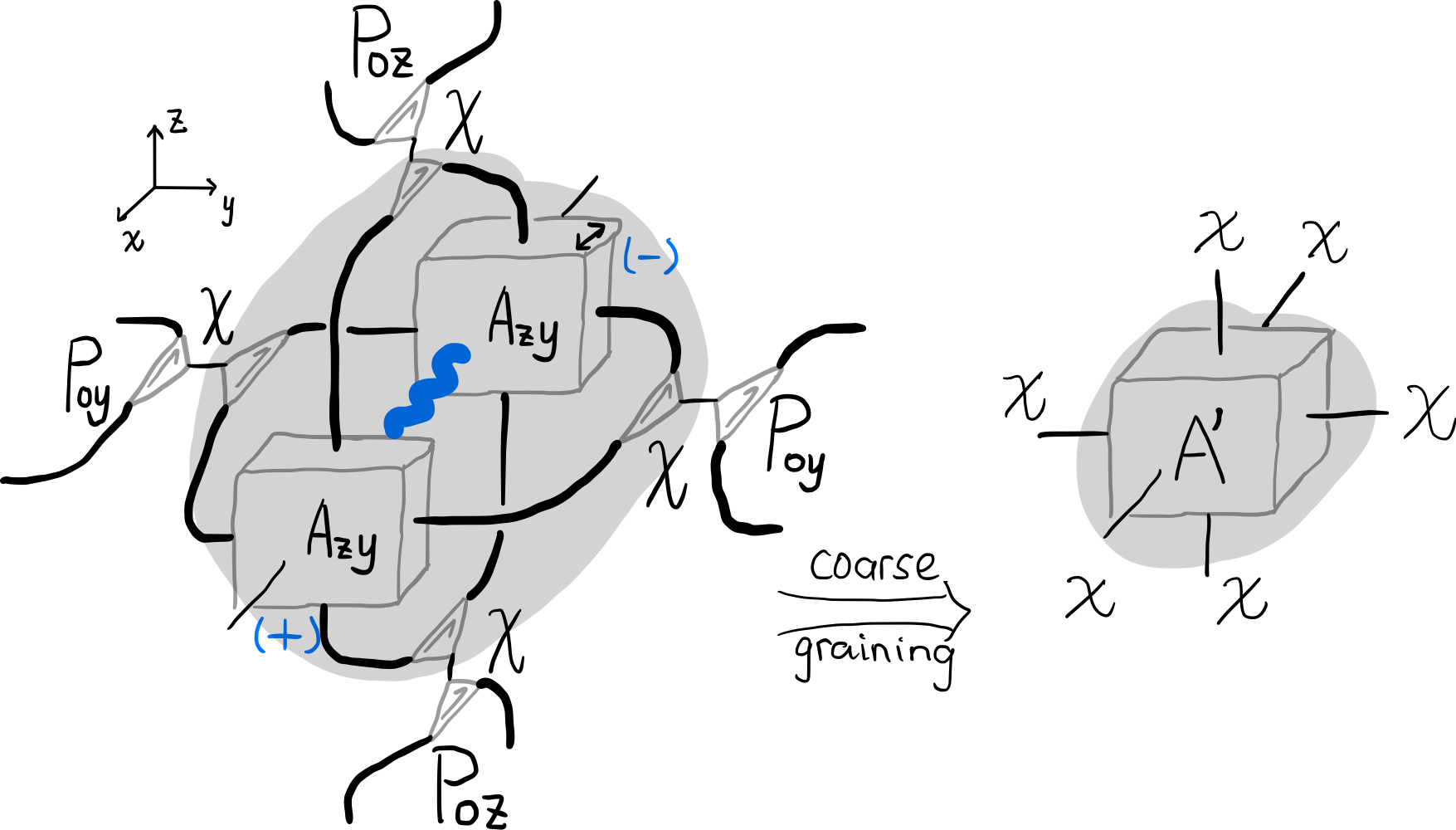}\,.
\end{align}
%{\color{orange}
    This map $A_{zy} \mapsto A'$ is the fourth and the final step of the proposed RG transformation.
%}
\end{subequations}
The computational costs of our block-tensor map are $O(\chi^{12.5})$ when we use $\chi_i = \chi^{1.5}, \chi_{ii} = \chi^2$ and $\chi_s \leq \chi_m \leq \chi$.
The composition of the entanglement filtering map, $A \mapsto A_s$ in Eq.~\eqref{eq:filtering}, with the HOTRG-like block-tensor transformation, $A_s \mapsto A_z \mapsto A_{zy} \mapsto A'$ in Eq.~\eqref{eq:hotrglikeBkten}, gives the proposed EF-enhanced block-tensor map $A \mapsto A'$.

% Benchmark
\section{An example}

\begin{figure*}[tb]
    \includegraphics[width=1.5\columnwidth,
    valign=c]{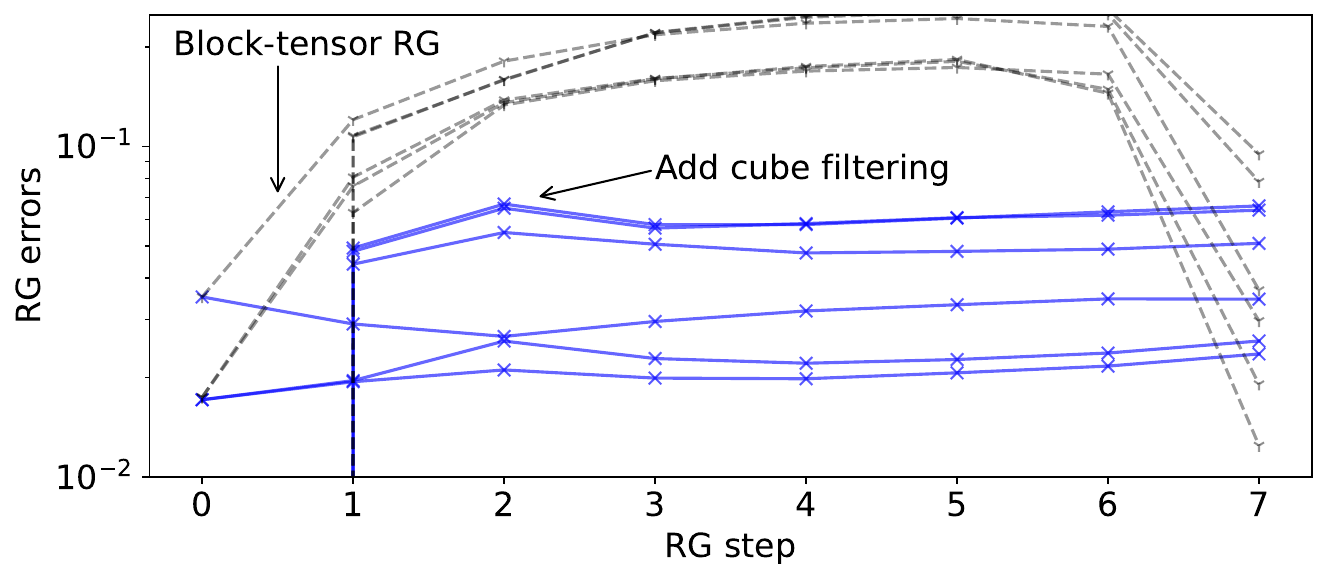}
    \includegraphics[width=1.5\columnwidth,
    valign=c]{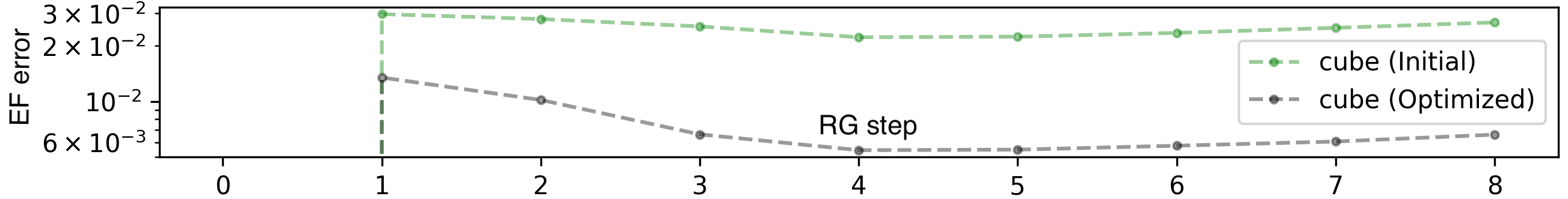}
    \caption{\label{fig:flowRGerrors}
        %{\color{orange}
        (Upper panel) The flow of RG errors of the go-to block-tensor scheme in 3D (HOTRG with $\chi=6$) are plotted as dashed lines.
        For the proposed method enhanced by cube filtering with $\chi=6, \chi_s = \chi_m=4$, we plot RG errors as solid lines. 
        There are multiple curves of the RG errors for both methods, since each RG step contains several projective truncations.
        (Lower panel) The cube filtering error (the ``EF error'' in the figure) in the proposed method.
        After the optimization, they should always be smaller than the RG errors.
        %}
}
\end{figure*}

We use the cubic-lattice Ising model to test the efficiency of the proposed RG transformation.
The linear growth of the entanglement entropy entails the growth of RG errors for a simple block-tensor scheme in 3D.
An effective EF scheme should be able to clear up the area-law term of the entanglement entropy, and thus keep the RG errors from growing with the RG step~\cite{Lyu:2023}.
To check this, we estimate the critical temperature $T_c$ using a shooting method~\cite{Lyu:2023,Ebel:2024Rot} (we briefly describe how to determine $T_c$ in Appendix~\ref{sec:app:findTc}) and plot the flow of the RG errors, as well as the errors of the cube filtering at the estimated $T_c$.
There are six projective truncations happening in a single RG step, and we plot them separately in~\autoref{fig:flowRGerrors} as solid lines.
For $\chi=6$ and $\chi_s=\chi_m=4$, the RG errors range from $2 \%$ to $6 \%$, while the cube filtering error is $0.6 \%$, one order of magnitude smaller than the RG errors.
Compared with the RG errors using the HOTRG (see the dashed lines in the upper panel in~\autoref{fig:flowRGerrors}), our proposed cube filtering successfully tames the growth of the RG errors in 3D.
Near the critical fixed point, the RG errors are reduced from more than $20\%$ to about $6\%$.
We want to emphasize here that the RG errors near the critical “fixed” point using the HOTRG will grow to more than $30\%$ when the bond dimension $\chi$ increases from 4 to 22. 
After adding the cube filtering, the maximum of all 6 RG errors near the critical fixed point decreases slowly when the bond dimension $\chi$ increases; it is $6\%, 7\%, 4\%, 2\%$ for $\chi = 6, 8, 11, 14$ (see the choice of hyperparameters in~\autoref{tab:xErrchi} and how to choose them in Appendix~\ref{sec:app:hyperpara}).

\par
\begin{figure*}[tb]
    \subfloat[Just block-tensor RG for $\chi=6$\label{fig:x6scD}]{
    \includegraphics[width=1.70\columnwidth,
    valign=c]{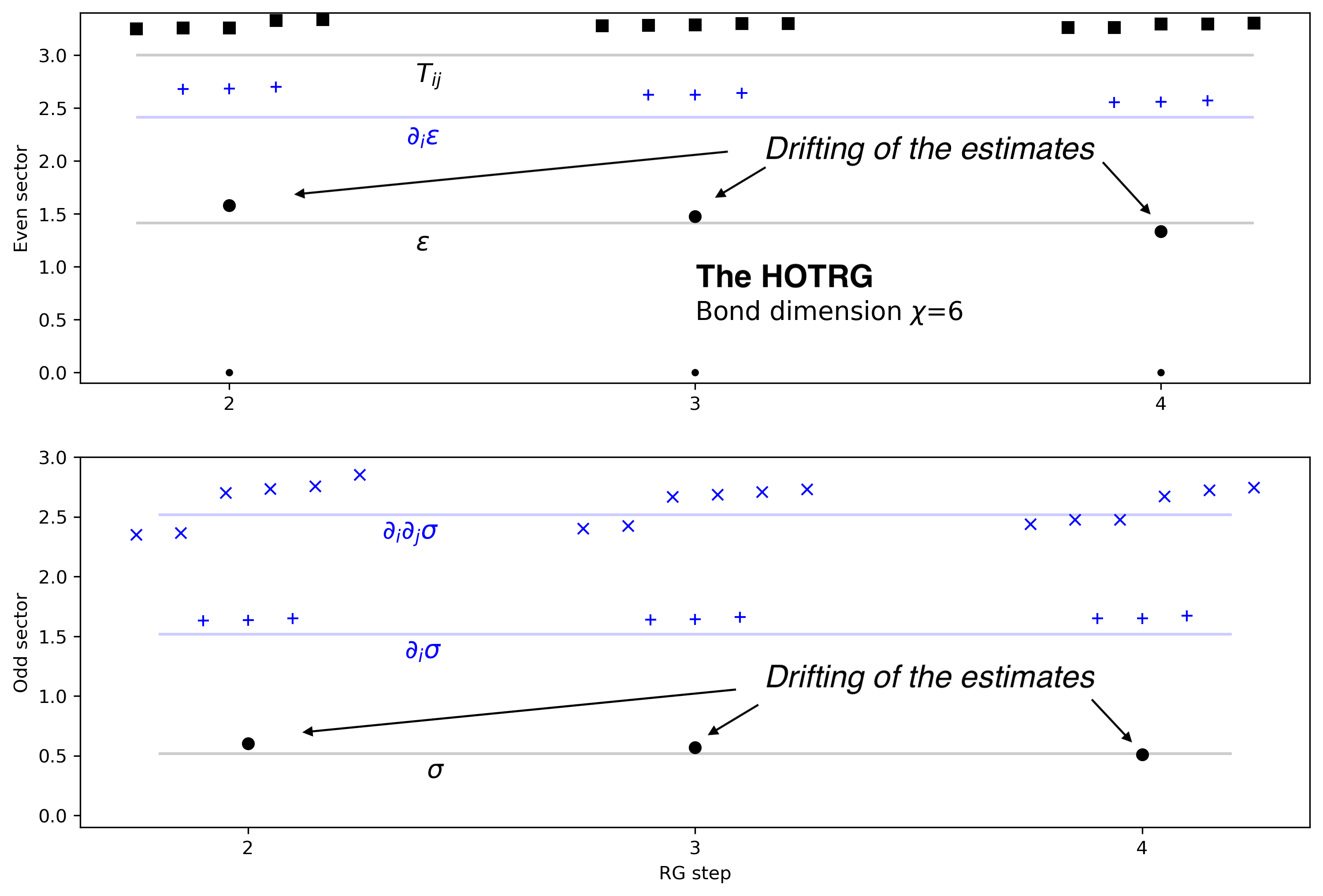}
}\\
\subfloat[With cube filtering for $\chi=6, \chi_s=\chi_m=4$\label{fig:x6s4m4scD}]{
    \includegraphics[width=1.70\columnwidth,
    valign=c]{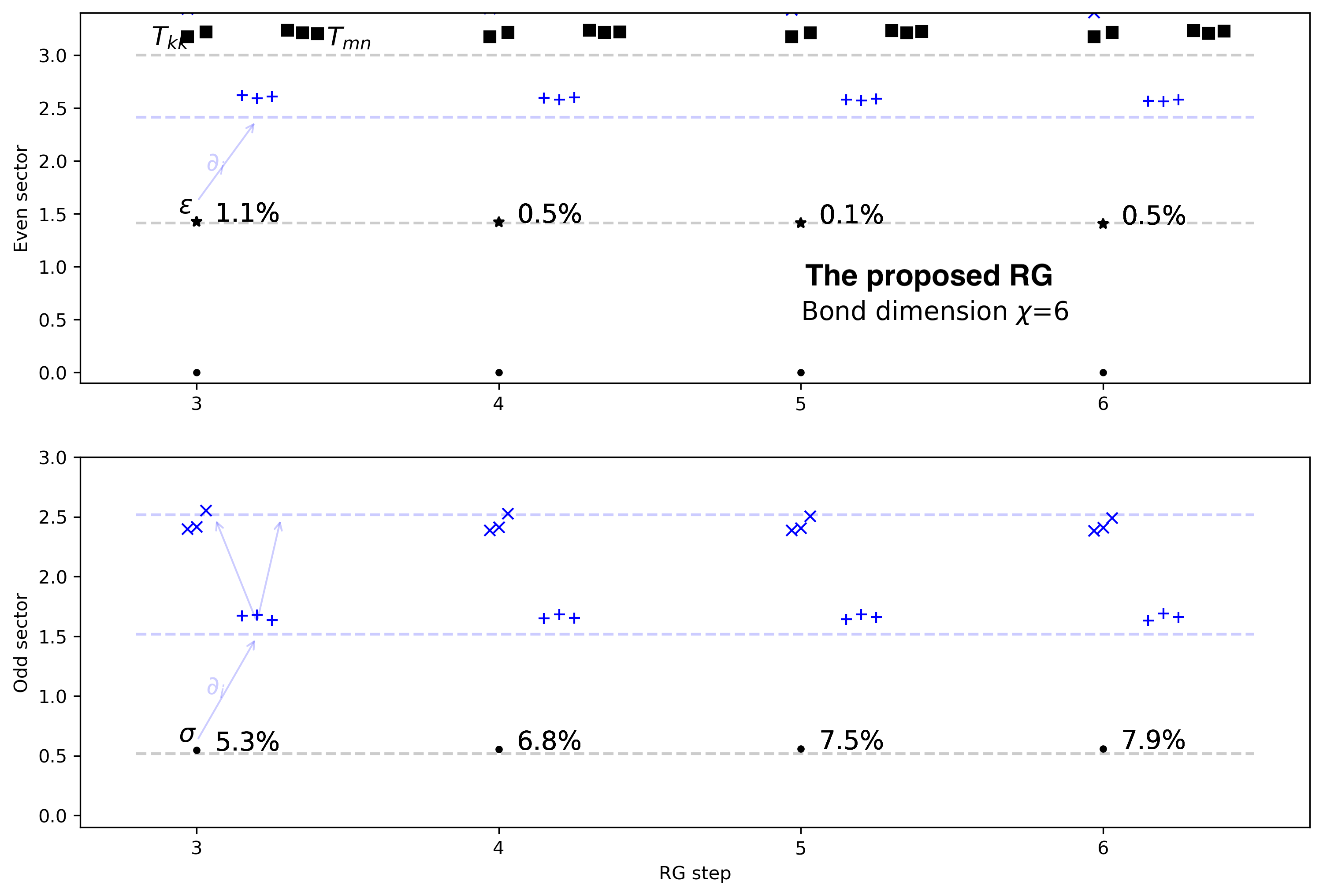}
}
    \caption{\label{fig:estScaleD}
        Estimates of scaling dimensions using a simple block-tensor scheme, the HOTRG, and the proposed RG, which is enhanced by entanglement filtering. 
        %{\color{orange}
            Horizontal dashed lines indicate the estimates by the conformal bootstrap.
            In both the HOTRG and the proposed RG, the spin-flip $\mathbb{Z}_2$ is exploited to extract scaling dimensions in spin-flip even and odd sectors separately.
            In the proposed RG, lattice reflection symmetry is also exploited, which gives another three symmetry charges that are not shown explicitly in the figure.
            We show scaling dimensions of the energy-density operator $\epsilon$, spin operator $\sigma$, and energy-momentum tensor $T_{ij}$, as well as some of their descendants like $\partial_i \epsilon, \partial_i \sigma$, etc.
            In the proposed RG, $T_{kk}$ and  $T_{mn}$ with $m \neq n$ can be distinguished by their lattice reflection symmetry charge.
            Next to the estimated scaling dimensions of $\epsilon$ and $\sigma$, the relative errors compared with the conformal bootstrap results are shown.
        %}
}
\end{figure*}

At the critical fixed point, we can linearize the RG map, from whose eigenvalue spectrum the scaling dimensions can be estimated~\cite{Lyu:2021,Ebel:2024LDO}.
A salient limitation of a 3D block-tensor map like the HOTRG is that the failure of exhibiting a critical fixed-point tensor~\cite{Lyu:2023}.
Due to this failure, the estimate of $x_{\epsilon}$ drifts with the RG step for most bond dimensions $\chi \leq 22$ (see Figure~\ref{fig:x6scD}, where $x_{\sigma}$ also drifts).
Adding the cube filtering allow the RG map to exhibit a critical fixed-point tensor;
the norm of the difference between two tensors of neighbor RG steps can in general decay down to order $10^{-2}$ (see~\autoref{fig:flowTenDiff}).
Moreover, the estimates of scaling dimensions converge with the RG step.
In Figure~\ref{fig:x6s4m4scD}, the percentages next to the estimates of $x_{\sigma}$ and $x_{\epsilon}$ are relative errors compared with the conformal bootstrap results.
At $\chi = 6, \chi_s=\chi_m=4$, the best estimate has relative error $0.1\%$ and $5.3\%$ for $x_{\epsilon}$ and $x_{\sigma}$.
We still have not observed a clear improvement of the estimates of $x_{\sigma},x_{\epsilon}$ when $\chi$ increases to 14.
The relative errors of the estimates for different $\chi$ are summarized in~\autoref{tab:xErrchi}.
The numerical results can be reproduced using the {\tt python} codes published at Ref.~\cite{Lyu:efrg3D}.

\begin{table}[htb]
    \caption{The estimation errors of $x_{\sigma}$ and $x_{\epsilon}$ versus the bond dimension $\chi$ and hyperparameters $\chi_s, \chi_m$ (compared with the conformal bootstrap results~\cite{Kos:Poland:Simmons:2016,Chang:Dommes:Erramilli:2024}).}\label{tab:xErrchi}
    \begin{ruledtabular}
    \begin{tabular}{ccccc}
        $\chi,\chi_s,\chi_m$             & 6,4,4   & 8,5,8   & 11,6,8  & 14,6,10  \\
        \colrule
        $x_{\sigma}$     & 5 $\sim$ 8\%   & 4 $\sim$ 6\% & 3 $\sim$ 6\% & 0.4 $\sim$ 0.5\% \\
        $x_{\epsilon}$     & 0.1 $\sim$ 1\% & 4 $\sim$ 5\% & 1 $\sim$ 6\% & 2 $\sim$ 4\% \\
    \end{tabular}
    \end{ruledtabular}
\end{table}

% Discussion
\section{Summary and perspective}
%{\color{orange}
    In this paper, we propose a 3D real space RG transformation using a tensor-network reformulation of Kadanoff's block idea, enhanced with an entanglement filtering procedure.
    The proposed RG is able to produce a fixed-point tensor near the critical fix point of the 3D Ising model.
    The RG approximation errors are stable with respect to the RG step near the critical fixed point, and they decrease from $6\%$ to $2\%$ when more couplings are retained in the RG map (when the bond dimension $\chi$ increases from $6$ to $14$).
    The relative errors of the estimated scaling dimensions of spin and energy-density operators, $x_{\sigma}, x_{\epsilon}$, are about $5\%$ for most bond dimensions, while there is no clear improvement when $\chi$ increases from $6$ to $14$.
    The best estimate of $x_{\sigma}$ has a relative error $0.4\%$ at a magical bond dimension $\chi=14$, and that of $x_{\epsilon}$ has a relative error $0.1\%$ at a magical bond dimension $\chi=6$.
    In summary, the proposed real space RG has well-controlled approximations, resolving the problems of uncontrolled approximations in spin-based RG transformations and of the growth of RG approximation errors in the 3D TNRG schemes using the simple block-tensor idea.
%}

Compared with existing techniques for analyzing critical systems, including Monte Carlo, series expansion and field theory methods, the TNRG can extract much more information.
A critical fixed-point tensor, if a TNRG scheme can produce, contains a complete description of a universality class.
The higher scaling dimensions in~\autoref{fig:estScaleD} exhibit conformal tower structure in 3D, which has never been reported numerically until quite recently in a fuzzy sphere construction~\cite{Zhu:Han:2023}.
Besides the scaling dimensions, operator product expansion coefficients can also be extracted from the fixed-point tensor~\cite{Guo:Wei:2024}.
Moreover, the method offers a bridge for connecting the RG theory with concepts in conformal field theory.
Recently, the relationship between the linearized RG and the dilatation operator has been clarified~\cite{Ebel:2024LDO}.
It also paves a way for an exact RG treatment of 3D criticality~\cite{Kennedy:Rychkov:2022,Kennedy:Rychkov:2023,Ebel:2024} and provides a playground for studying general properties of RG, such as the entropic c-theorems~\cite{zomo-ctheorem:1986,Casini:2007,Casini:2015,Nishioka:2018}.
Due to the high computational costs in 3D, the reachable bond dimension is small, making it harder to demonstrate whether the proposed RG is systematically improvable.
The ideas for reducing the computational costs of the 3D TNRG~\cite{Adachi:2020,Kadoh:2019,Kadoh:2022} might make the proposed RG a more powerful tool for cracking unsolved problems in 3D classical or (2+1)D quantum criticality.

% Acknowledgments starting here
\acknowledgments

%{\color{orange}
    We thank Slava Rychkov for comments about the role of lattice reflection symmetry in the proposed algorithm.
%}
X.L.\ is grateful to the support of the Global Science Graduate Course (GSGC) program of the University of Tokyo.
This work is financially supported by MEXT Grant-in-Aid for Scientific Research (B) (23H01092).
The computation in this work has been done using the facilities of the Supercomputer Center, the Institute for Solid State Physics, the University of Tokyo.
Most part of the manuscript was written after X.L.\ moved to Institut des Hautes \'Etudes Scientifiques.

% Appendices starting here
\appendix
%{\color{orange} 
\section{About the lattice reflection symmetry}\label{sec:app:reflsym}
In the proposed RG transformation, the transposition trick is used to exploit the lattice reflection symmetry of the underlying model.
We will expound how to exploit lattice reflection symmetry in a TNRG setting in a coming paper.
At this point, we want to point out that the transposition trick in the proposed method is applicable to models like ferromagnetic Ising model and $n$-state Potts model on square and cubic lattice with uniform nearest-neighbor couplings.
The reason is that one can show that the transposition trick keeps the partition function invariant.

For models that do not have lattice reflection symmetry, here is how to turn off the transposition trick. 
The same tensor $A$ is put in the $8$ vertices of the cube in Eq.~\eqref{eq:transposeTrick} without transpositions.
When inserting the filtering matrices in the cube (see Eq.~\eqref{eq:EF3Dapprox}), the two matrices on a given bond are different, and matrices on different bonds are not related.
In total, there are $12 \times 2 = 24$ filtering matrices in Eq.~\eqref{eq:EF3Dapprox} for the $12$ inner bonds of the cube.
When these 24 filtering matrices acts on the tensor $A$ in the TNRG $2 \times 2 \times 2$ block (see Eq.~\eqref{eq:filtering}), each $A$ among the eight vertices absorbs three filtering matrices through its three outer legs according to the position of the tensor $A$ in the TNRG block.
After the absorption of these filtering matrices, the TNRG block in Eq.~\eqref{eq:filtering} has eight different tensors, which are no longer related to each other through transpositions.
A similar HOTRG-like block-tensor map can be applied to this TNRG block after the filtering.

Now, the numerical advantage of exploiting the lattice reflection symmetry becomes clear.
The number of independent filtering matrices reduces from $24$ to $3$.
This simplifies the process of determining the filtering matrices, how these filtering matrices acts on the tensor $A$, and the HOTRG-like block-tensor process.
%}

%{\color{orange} 
\section{Estimation of the critical temperature and the tensor RG flows at criticality}\label{sec:app:findTc}
\begin{figure*}[htb]
    \subfloat[Just block-tensor RG (the HOTRG) for $\chi=8$]{
    \includegraphics[width=0.95\columnwidth,
    valign=left]{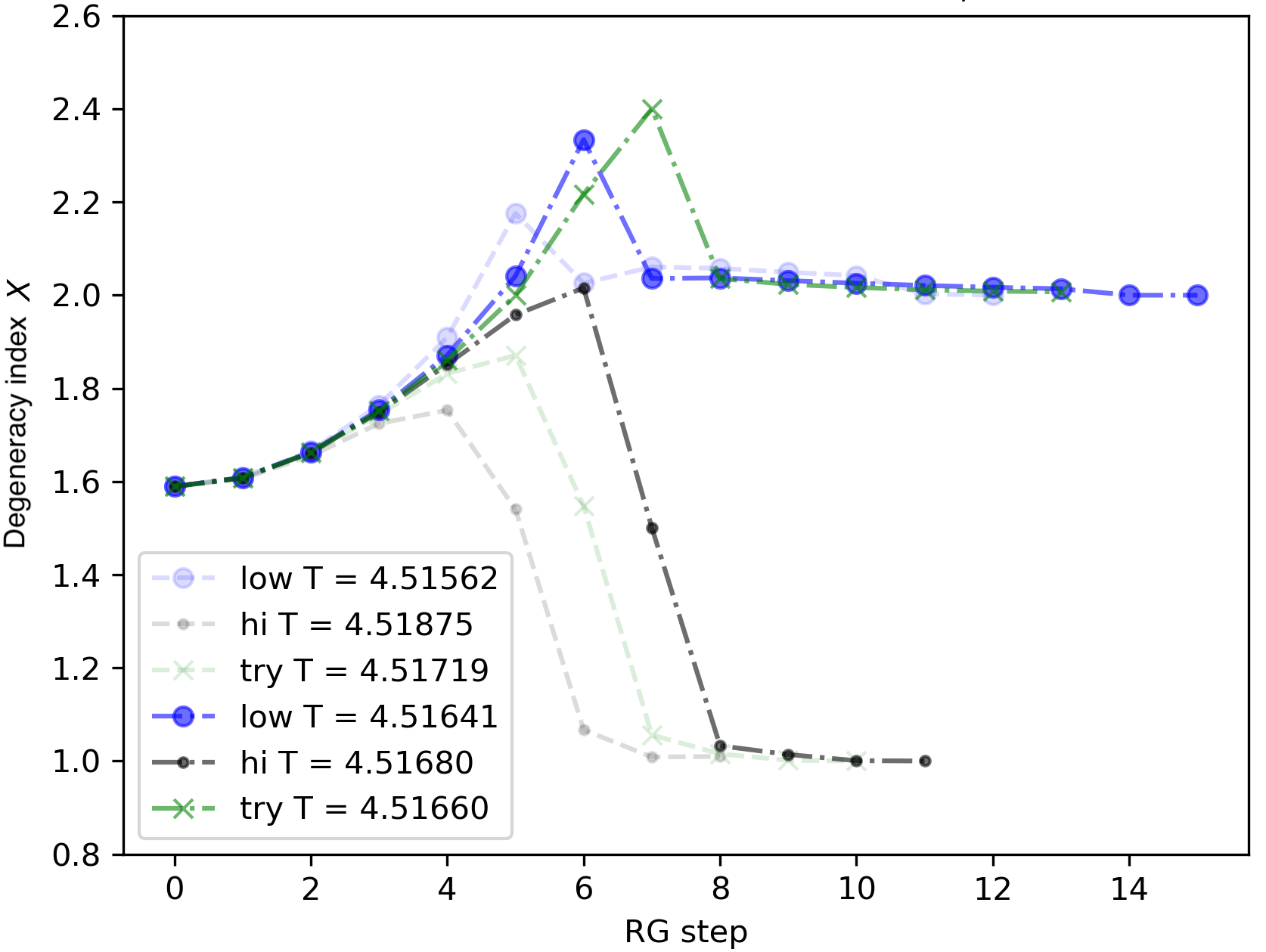}
}
    \subfloat[With cube filtering for $\chi=8, \chi_s=5, \chi_m=8$]{
    \includegraphics[width=0.95\columnwidth,
    valign=right]{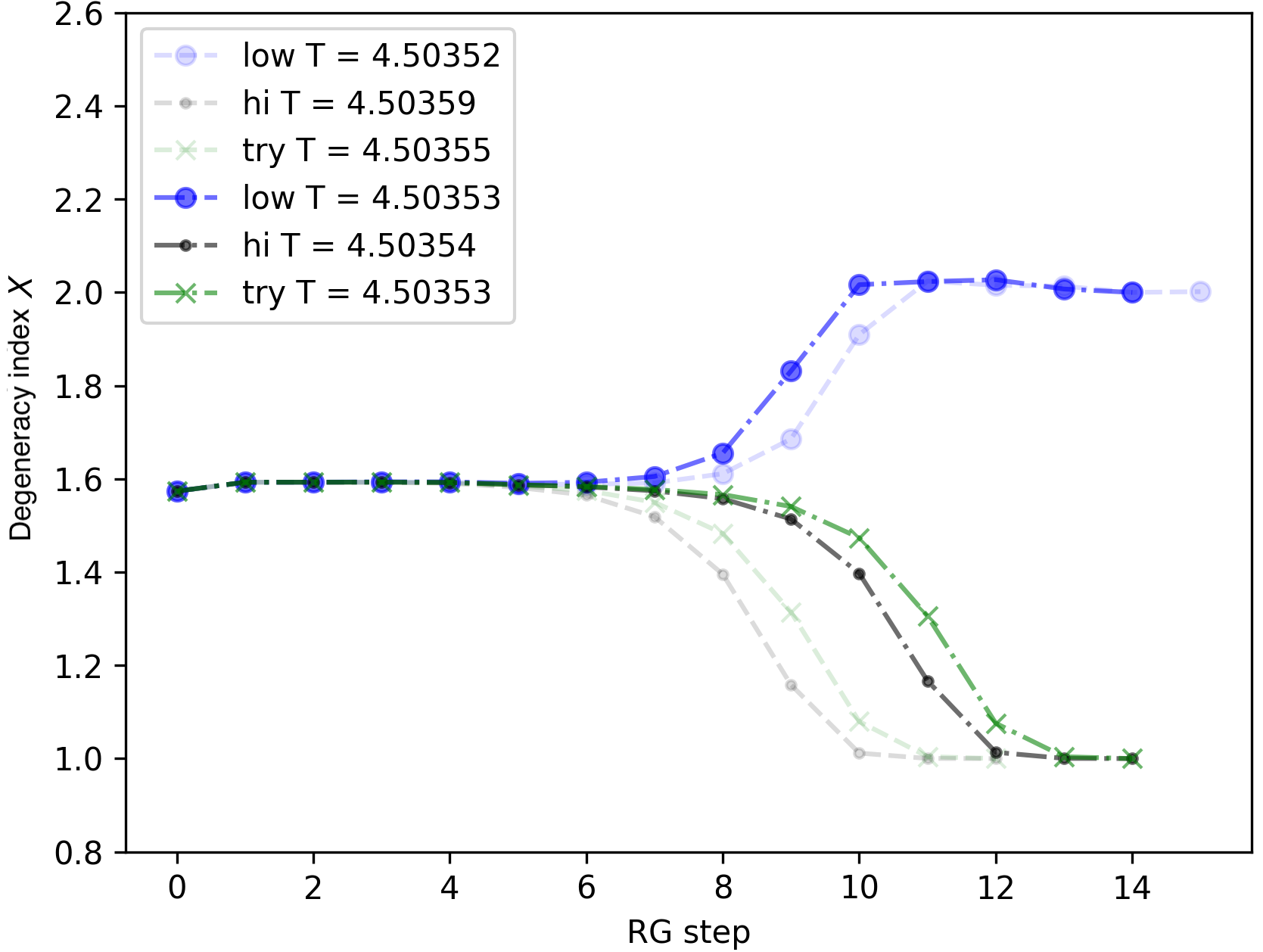}
}
    \caption{\label{fig:flowDegInd}
        RG flow of the degeneracy index $X$ before and after entanglement filtering
}
\end{figure*}

\begin{figure*}[htb]
    \subfloat[Just block-tensor RG (the HOTRG) for $\chi=8$. Without entanglement filtering, the difference only decreases to about $5\times 10^{-2}$ until RG step $n=5$, after which the difference starts to grow.]{
    \includegraphics[width=0.68\textwidth,
    valign=left]{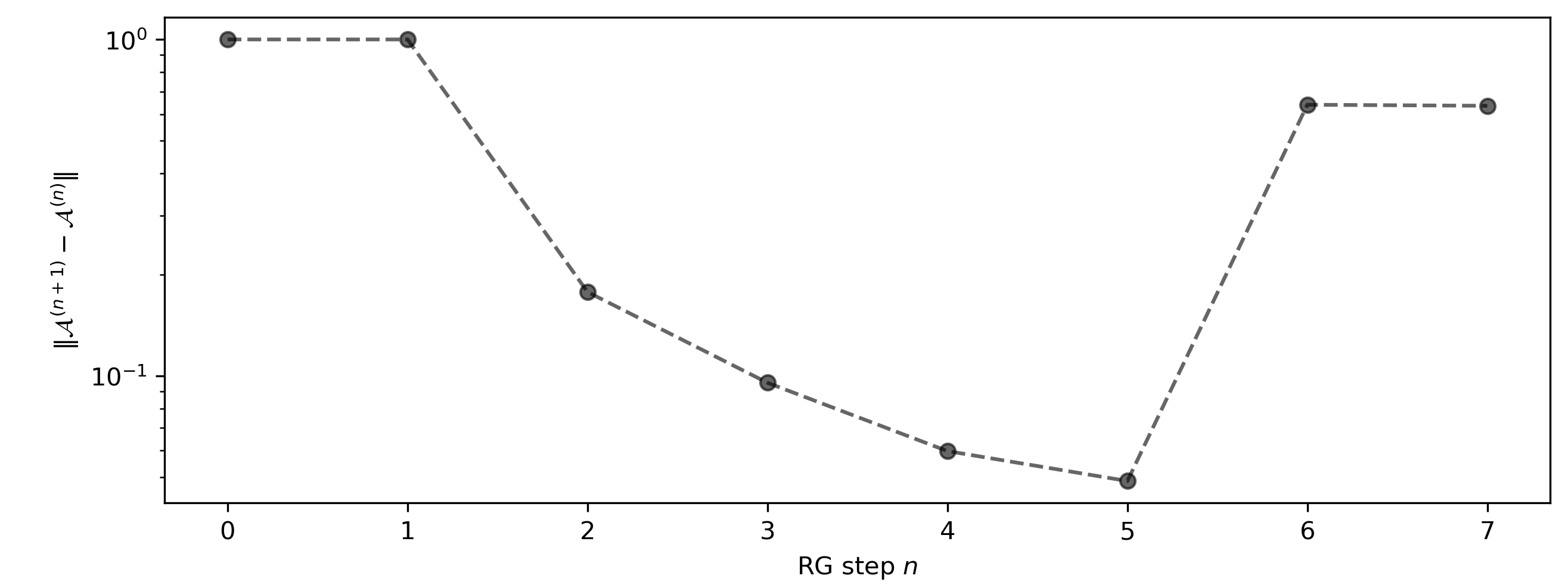}
}

\subfloat[With cube filtering for $\chi=8, \chi_s=5, \chi_m=8$. The difference decreases to about $4\times 10^{-3}$, one order of magnitude smaller than the simple block-tensor RG.]{
    \includegraphics[width=0.68\textwidth,
    valign=right]{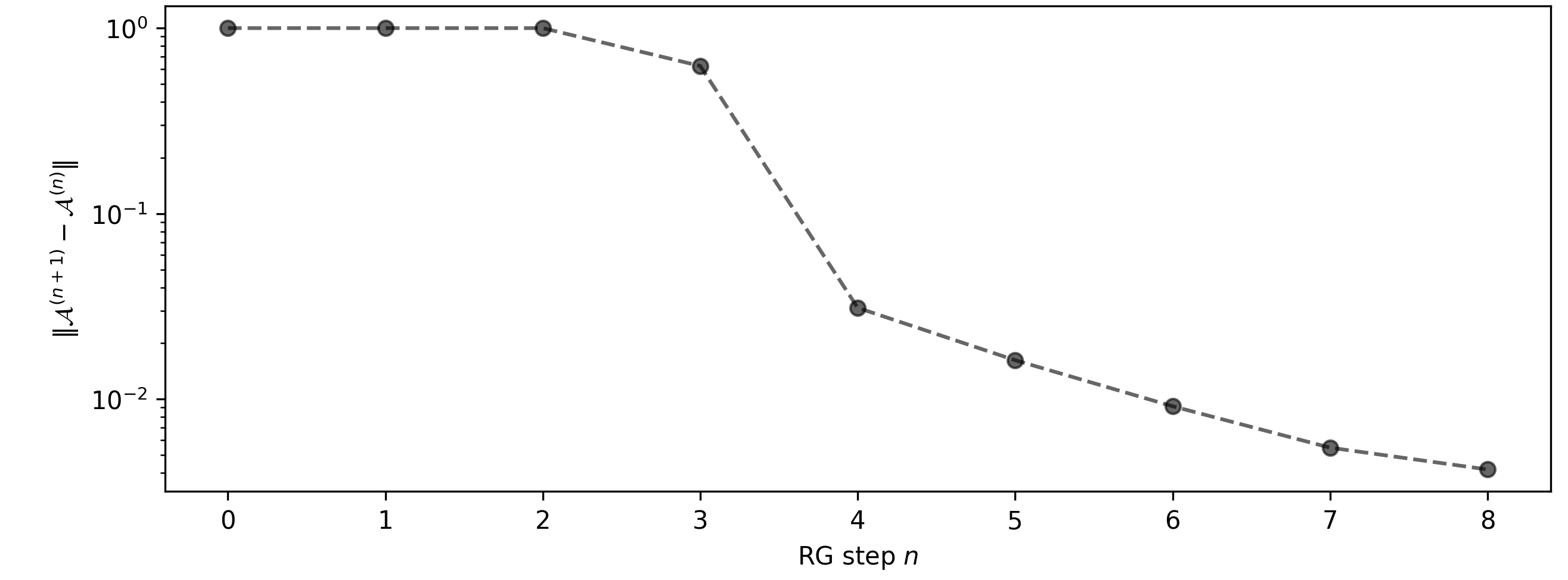}
}
    \caption{\label{fig:flowTenDiff}
        For the tensor RG flow generated at the estimated critical temperature, we plot the Frobenius norm of the difference between two tensors of adjacent RG steps: $||\mathcal{A}^{(n+1)} - \mathcal{A}^{(n)}||$, where $n$ denotes the RG step.
        The tensor is normalized to have unit Frobenius norm $||\mathcal{A}^{(n)}|| = 1$.
        Furthermore, gauge fixing~\cite{Lyu:2021} is necessary and has been done here.
}
\end{figure*}

The critical temperature can be estimated by analyzing the tensor RG flows.
Starting with a temperature, say $T=4.48$, we generate an RG flow in the tensor space.
An easy way to find out the phase that a tensor corresponds to is defining the following degeneracy index $X$,
\begin{align}
    \label{eq:defDegInd}
    \includegraphics[scale=0.05, valign=c]{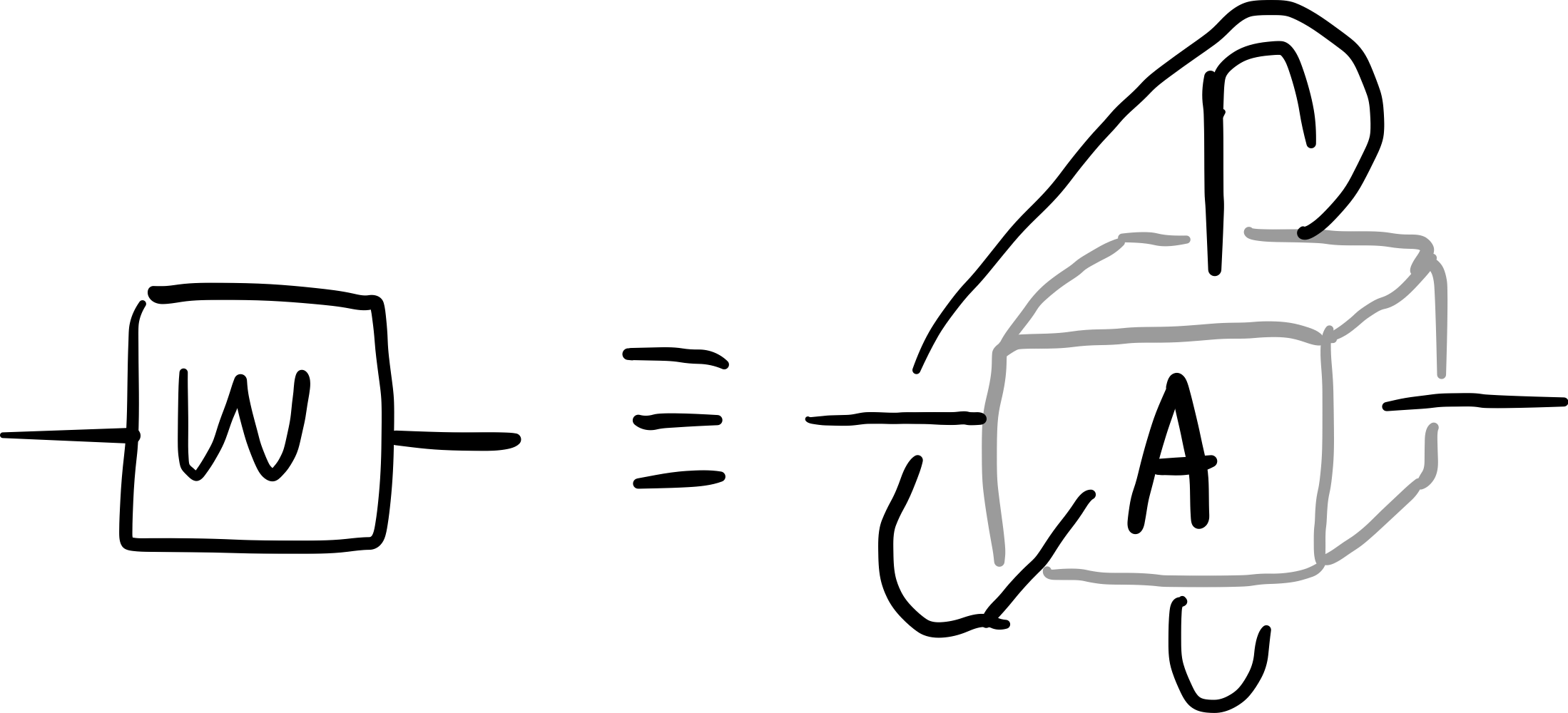}\quad\text{and}\quad
    X \equiv \frac{(\tr{W})^2}{\tr{W^2}},
\end{align}
which reduces a tensor to a single number.
The high-$T$ fixed point has $X = 1$ and the low-$T$ fixed point has $X = 2$.
If, for example, the initial tensor at temperature $T = 4.48$ flows to the low-$T$ fixed point, we know that the lower bound of the critical temperature is $T_{\text{low}} = 4.48$.
Afterward, we choose a higher temperature, $T = 4.54$, for example, so that the initial tensor flows to the high-$T$ fixed point; hence, we know the upper bound of the critical temperature is $T_{\text{hi}} = 4.54$.
Using bisection method, we can estimate the critical temperature of the model to a high precision so that the tensor stays stable near the critical fixed point for a while.
\autoref{fig:flowDegInd} demonstrates that the entanglement filtering process is essential for producing a critical fixed-point tensor that is actually fixed.
The block-tensor transformation fails to do so in 3D due to the linear growth of the entanglement entropy~\cite{Lyu:2023}.

To demonstrate the convergence property of the proposed RG map near the critical fixed point, we generate the tensor RG flow at the estimated critical temperature, perform the gauge fixing according to Ref.~\cite{Lyu:2021}, and plot the Frobenius norm of the difference between two tensors of adjacent RG steps: $||\mathcal{A}^{(n+1)} - \mathcal{A}^{(n)}||$ (see~\autoref{fig:flowTenDiff}).
The tensor $\mathcal{A}^{(n)}$ here is the normalized version of the tensor $A^{(n)}$ in Eq.~\eqref{eq:tenRGflow} with unit Frobenius norm $||\mathcal{A}^{(n)}|| = 1$.
After incorporating the cube filtering, this difference decreases for more RG steps to about $4 \times 10^{-3}$, which is one order of magnitude smaller than the simple block-tensor RG case.

%}

%{\color{orange}
    \section{Choice of the hyperparameters in the proposed RG}\label{sec:app:hyperpara}
Apart from the bond dimension $\chi$, the proposed RG map has a few more parameters related to it.
In the cube filtering, we squeeze $\chi$ to a smaller one $\chi_s < \chi$, which introduces one more parameter $\chi_s$.
The computational costs of determining the filtering matrices are $O(\chi^{12})$, which comes from the construction of the bond environment matrix for the bond to be filtered.
In the HOTRG-like block-tensor map, we make the bond dimension of the intermediate outer legs a tunable parameter $\chi_{m}$, which is chosen to be in the range $\chi_s \leq \chi_m \leq \chi$.
\emph{
    These two addition parameters $\chi_s,\chi_m$ should be tuned in such a way that the errors in the HOTRG-like block-tensor map is reduced compared with the plain HOTRG without entanglement filtering, and remain stable with respect to the RG step.
In the meantime, the approximation error in the cube filtering should be smaller than errors in the block-tensor map.
(see~\autoref{fig:flowRGerrors})
}
The two bond dimension of the inner legs in the HOTRG-like block-tensor transformation is chosen to make sure the projective truncation errors of inner legs are smaller than those of outer legs.
Our numerical experiments using the 3D Ising model suggestions the following rule of thumb: $\chi_i = \chi^{1.5}$ and $\chi_{ii} = \chi^{2}$.
With this choice, there are only two hyperparameters $\chi_s, \chi_m$ in the proposed method for a given bond dimension $\chi$.
The computational costs of the HOTRG-like block-tensor map are $O(\chi^{12.5})$, where the computationally heaviest part is the $y$ collapse; thus, the proposed scheme has computational costs $O(\chi^{12.5})$.
%}

% The below uses bibtex to construct the bibliography.

% \bibliographystyle{apsrev4-2}

\bibliography{references}     % This is a reference to "references.bib".

\end{document}